\documentclass{techrep}
\usepackage{CFR}

\usepackage{nicefrac}
\usepackage{amsmath,amssymb}

\usepackage{fancyhdr}
\fancypagestyle{ack}{%
  \fancyhf{}

  \fancyfoot[L]{{\small This is a preprint of a chapter forthcoming in \emph{Handbook of Mixture Analysis},
    edited by Gilles Celeux, Sylvia Fr{\" u}hwirth-Schnatter, and Christian P. Robert.}}
}

\newcommand{\Randi}{r_a}

\newcommand{\tr}{\mbox{\rm tr}}
\newcommand{\diag}{\mbox{\rm Diag}}
\newcommand{\Omath}{\mathcal{O}}

\author{Gilles Celeux\\INRIA Saclay, France \And 
  Kaniav Kamary\\INRIA Saclay, France \AND
  Gertraud Malsiner-Walli\\Vienna University of Economics and\\Business, Austria \And
  Jean-Michel Marin\\Universit\'e de Montpellier, France \AND
  Christian P. Robert\\Universit\'e Paris-Dauphine, France and\\University of Warwick, UK}

\title{Computational Solutions for Bayesian Inference in Mixture Models}

\Plainauthor{G. Celeux, K. Kamary, G. Malsiner-Walli, J.-M. Marin, and C. P. Robert}

\begin{document}
\maketitle
\thispagestyle{ack}




\vspace{-0.8cm}
\tableofcontents
\vspace{0.4cm}

This chapter surveys the most standard Monte Carlo methods available for simulating from a posterior distribution associated with a mixture and conducts some experiments about the robustness of the Gibbs sampler in high dimensional Gaussian settings.
\section{Introduction}

It may sound paradoxical that a statistical model that writes as a sum of
Gaussian densities poses a significant computational challenge to Bayesian (and
non-Bayesian alike) statisticians, but it is nonetheless the case.  Estimating
(and more globally running a Bayesian\index{Bayesian!inference|(} analysis on)
the parameters of a mixture model has long been deemed impossible, except for
the most basic cases, as illustrated by the approximations found in the
literature of the 1980's
\citep{smith:makov:1978,titterington:smith:makov:1985,bernardo:giron:1988,crawford:degroot:kadane:small:1992}. Before
the introduction of Markov chain Monte Carlo (MCMC)\index{MCMC} methods to the Bayesian community,
there was no satisfying way to bypass this problem and it is no surprise that
mixture models were among the first applications of Gibbs sampling to appear
\citep{diebolt:robert:1990a,gelman:king:1990,west:1992}.  The reason for this
computational challenge is the combinatoric explosive nature of the development
of the likelihood function, which contains $G^n$ terms when using $G$
components over $n$ observations. As detailed in other chapters, like
Chapter~1, the natural interpretation of the likelihood is to
contemplate all possible partitions of the $n$-observation sample $\ym=(y_1,
\ldots, y_n$). While the likelihood function can be computed in
$ \Omath (G \times n)$ time, being expressed as
$$
p (\ym| \theta) = \prod_{i=1}^n \sum_{g=1}^G \eta_g f(y_i|\theta_g)\,,
$$
where $\theta=(\theta_1,\ldots,\theta_G,\eta_1,\ldots,\eta_G)$, there is no
structure in this function that allows for its exploration in an efficient way. Indeed, as demonstrated for
instance by Chapter~2,  the variations of this function are not readily available and require completion steps as in
the EM algorithm. Given a specific value of $\theta$, one can compute $p(\ym| \theta)  $ but this numerical value
does not bring useful information on the shape of the likelihood in a neighbourhood of $\theta$. The value of the
gradient also is available in  $\Omath (G  \times n)$ time,  but does not help much in this regard.
(Several formulations of the Fisher information matrix for, e.g., Gaussian mixtures through special functions
are available, see, e.g., \citealp{behboodian:1972} and \citealp{cappe:moulines:ryden:2004}.)

Computational advances have thus been essential to Bayesian inference on
mixtures while this problem has retrospectively fed new advances in Bayesian
computational methodology.\footnote{To some extent, the same is true for the
pair made of maximum likelihood estimation and the EM algorithm, as discussed in Chapter~2.} In
Section~\ref{section2Alg}  we cover some of the proposed
solutions, from the original Data Augmentation\index{data augmentation} of
\cite{tanner:wong:1987} that predated the Gibbs\index{Gibbs sampling} sampler
of \cite{gelman:king:1990} and \cite{diebolt:robert:1990a} to specially
designed algorithms, to the subtleties of label switching
\citep{stephens:2000}.


As stressed by simulation experiments in Section~\ref{chaptersim}, there
nonetheless remain major difficulties in running Bayesian inference on mixtures
of moderate to large dimensions. First of all, among  all the algorithms that
will be reviewed  
only Gibbs sampling seems to scale to high dimensions. Second,
the impact of the prior distribution remains noticeable for sample sizes that
would seem high enough in most settings, while larger sample sizes see the
occurrence of extremely peaked posterior distributions that are a massive
challenge for exploratory algorithms like MCMC methods.
Section~\ref{secprihigh} is specifically devoted to  Gibbs sampling  for
high-dimensional Gaussian mixtures and a new prior distribution is introduced
 that seems to scale appropriately to high dimensions.

\section{Algorithms for Posterior Sampling}  \label{section2Alg}

\subsection{A computational problem? Which computational problem?}

\begin{figure}[t!]
    \centering
    \includegraphics[width=0.55\textwidth,angle=0]{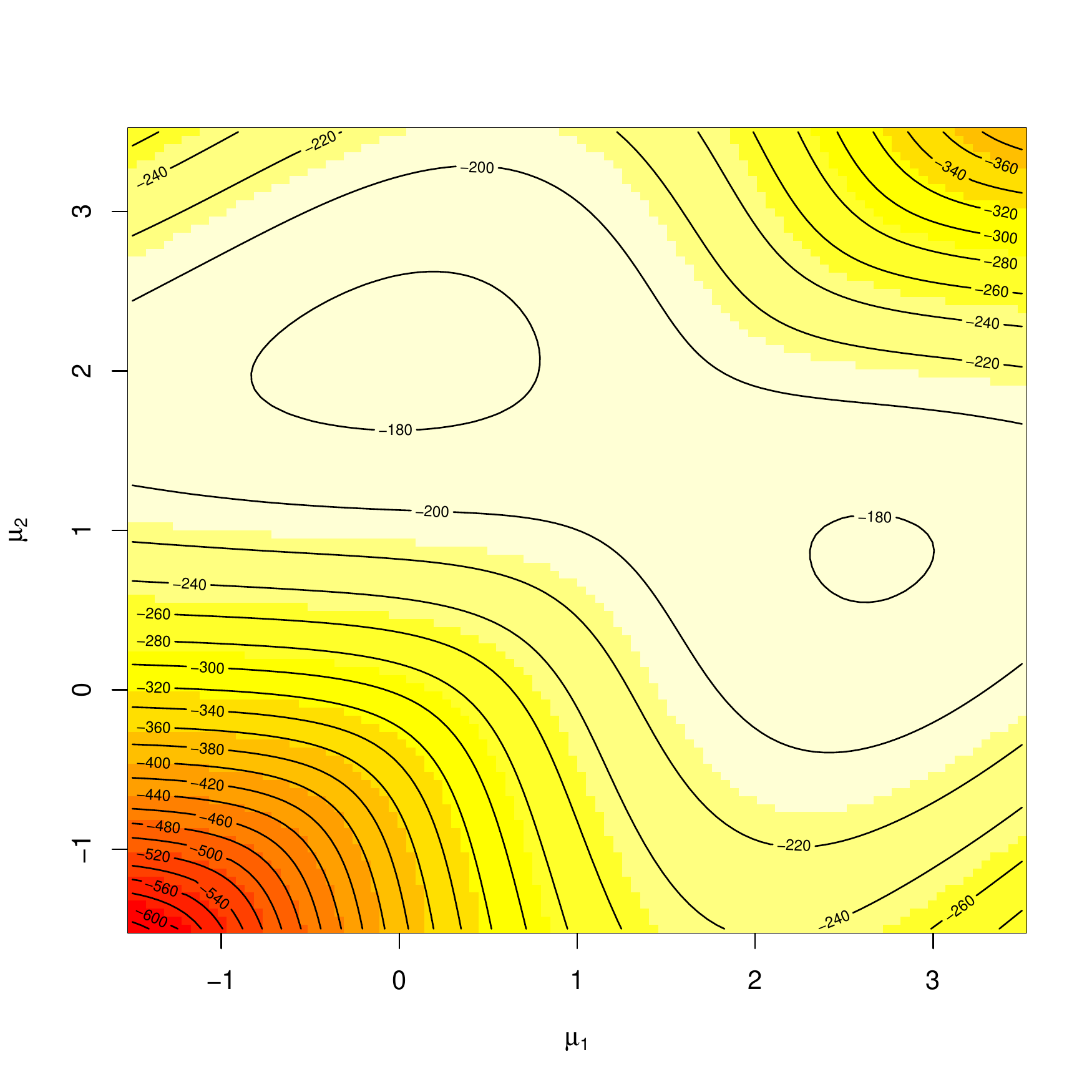}
\caption{\label{fig:trupost}
Posterior density surface for $100$ observations from the Gaussian mixture
$\nicefrac{3}{10}\mathcal{N}(\mu_1,1)+\nicefrac{7}{10}\mathcal{N}(\mu_2,1)$,
clearly identifying a modal region near the true value of the parameter $(\mu_1,\mu_2)=(0,\nicefrac{5}{2})$ used to
simulate the data, plus a secondary mode associated with the inversion of the data allocation to the two components. The figure was produced by computing the likelihood and prior functions over a $100\times100$ grid over $(-1,3)\times(-1,3)$. The level sets in the image are expressed in a logarithmic scale. }
\end{figure}

When considering a mixture model from a Bayesian perspective (see, e.g., Chapter~4),
the associated posterior distribution
$$
p(\theta|\by) \propto p(\theta)\prod_{i=1}^n \sum_{g=1}^G \eta_g f(y_i|\theta_g),
$$
based on   the prior distribution $ p(\theta)$,  is available in closed form, up to the normalising constant, because the number
of terms to compute is of order $\Omath (G\times n)$ in dimension one and of
order $\Omath (G\times n \times d^2)$ in dimension $d$. This means that two
different values of the parameter can be compared through their (non-normalized) posterior
values.
 See for instance Figure \ref{fig:trupost} that displays the posterior
density surface in the case of the univariate Gaussian mixture
$$
\frac{3}{10}\,\mathcal{N}(\mu_1,1)+\frac{7}{10}\,\mathcal{N}(\mu_2,1)\,,
$$
clearly identifying a modal region near the true value of the parameter
$(\mu_1,\mu_2)=\left(0,\frac{5}{2}\right)$ actually used to simulate the data.
However, this does not mean a probabilistic interpretation of $p(\theta|\by)$ is immediately manageable:
deriving posterior means, posterior variances, or simply identifying regions of high posterior density value remains a
major difficulty when considering only this function. Since the dimension of the parameter space grows quite rapidly with
$G$ and $d$, being for instance $3G-1$ for a unidimensional Gaussian mixture against $G-1+dG+(d(d+1)/2)G$ for a
$d$-dimensional Gaussian mixture,\footnote{This $\Omath (d^2)$ magnitude of the parameter space explains why we {\em in fine} deem
Bayesian inference for generic mixtures in large dimensions to present quite an  challenge.}
numerical integration\index{numerical integration} cannot be considered as a viable alternative.
Laplace\index{Laplace approximation}  approximations \citep[see, e.g.,][]{rue:martino:chopin:2009} are incompatible with the multimodal nature of the posterior distribution. The only practical solution
thus consists of producing a simulation technique that approximates outcomes from the posterior. In principle, since the
posterior density can be computed up to a constant, MCMC algorithms\index{MCMC} should operate smoothly in this setting.
As we will see in this chapter, this is not always the case.

\subsection{Gibbs sampling}  \label{sec5Gibbs}

Prior\index{Gibbs sampling|(} to 1989 or more exactly to the publication by \cite{tanner:wong:1987} of their Data Augmentation\index{data augmentation} paper, which can be considered as a precursor to the Gelfand and Smith \citeyear{gelfand:smith:1990} Gibbs sampling paper, there was no manageable
way to handle mixtures from a Bayesian perspective. As an illustration on an univariate Gaussian mixture with two components,
\cite{diebolt:robert:1990b} studied a \lq\lq grey\index{grey code}
code\rq\rq\  implementation\footnote{That is, an optimised algorithm that minimises computing costs.}
for exploring the collection of all partitions of a sample of size $n$ into $2$ groups and managed to reach $n=30$
within a week of computation (in 1990's standards).

Consider thus a mixture of $G$ components
\begin{equation}\label{eq:7.4}
\sum_{g=1}^G \; \eta_g \; f(y|\theta_g) \;,\qquad y\in\mathbb{R}^d\,,
\end{equation}
where, for simplicity's sake, $f(\cdot|\theta)$ belongs to an exponential family
$$
f(y|\theta) = h(y) \; \exp\{\theta\cdot y - \psi(\theta)\}
$$
over the set $\mathbb{R}^d\,,$ and where
$(\theta_1,\ldots,\theta_G)$ is distributed from a
product of conjugate priors
$$
p (\theta_g|\alpha_g,\lambda_g)\propto
\exp\{\lambda_g(\theta_g\cdot\alpha_g-\psi(\theta_g))\} \;,
$$
with hyperparameters $\lambda_g >0$ and $\alpha_g\in\mathbb{R}^d$ $(g=1,\ldots,G)$,
while $(\eta_1,\ldots,\eta_G)$ follows the usual Dirichlet prior:
$$
(\eta_1,\ldots,\eta_G) \sim \Dir (e_1,\ldots,e_G) \;.
$$
These prior choices are only made for convenience sake, with hyperparameters requiring inputs from the modeller. Most obviously,
alternative priors can be proposed, with a mere increase in computational complexity.
Given a sample $(y_1,\ldots,y_n)$ from (\ref{eq:7.4}), then  as already explained in Chapter~1,
we can associate to every observation an indicator random variable
$z_i \in \{1,\ldots,G\}$ that indicates which component of the mixture is associated with $y_i$, namely which term in
the mixture was used to generate $y_i$.  The demarginalization (or {\it completion}) of model (\ref{eq:7.4}) is then
$$
z_i \sim \Mulnom (1, \eta_1,\ldots,\eta_G), \qquad
y_i|z_i \sim f(y_i |\theta_{z_i})\;.
$$
Thus, considering $x_i = (y_i,z_i)$ (instead of $y_i$) entirely eliminates the mixture structure from the model
since the likelihood of the completed model (the so-called complete-data   likelihood function) is given by:
\begin{eqnarray*}
\likc (\theta|x_1,\ldots,x_n)
&\propto& \prod_{i=1}^n \; \eta_{z_i} \; f(y_i|\theta_{z_i}) \\
&=& \prod_{g=1}^G \; \prod_{i;z_i=g} \; \eta_g \; f(y_i|\theta_g) \;.
\end{eqnarray*}

\begin{algorithm}[t!]\caption{Mixture Posterior Simulation} \label{al:miy_sim}
Set hyperparameters $(\alpha_g,e_g,\lambda_g)_g$.
Repeat the following steps $T$ times, after a suitable burn-in phase:
 \begin{algorithmic}\itemsep 1mm
\item[{\tt 1.}] {\tt Simulate $z_i \;\; (i=1,\ldots,n)$ from}
$$
\Prob (z_i=g|\theta_g, \eta_g, y_i) \propto \eta_g \; f(y_i|\theta_g),  \qquad (g=1,\ldots,G),
$$
{\tt and compute the statistics}
$$
n_g = \sum_{i=1}^n \; \mathbb{I}(z_i=g) \;, \qquad
n_g {\overline y}_g = \sum_{i=1}^n \; \mathbb{I} (z_i=g) y_i \;.
$$
\item[{\tt 2.}] {\tt Generate} $(g=1,\ldots,G)$
$$
\theta_g  |\zm, \ym \sim \displaystyle{p \left(\theta_g \bigg|{\lambda_g \alpha_g+n_g
{\overline y}_g \over \lambda_g+n_g},\lambda_g+n_g\right) }\,,
$$
$$
(\eta_1,\ldots,\eta_G)  |\zm, \ym  \sim \Dir _G (e_1+n_1,\ldots,e_G+n_G) \;.
$$
\end{algorithmic}
\end{algorithm}

\noindent This latent structure is also exploited in the original implementation of the EM algorithm, as discussed in Chapter~2. Both steps of the Gibbs sampler \citep{robert:casella:2004} are
then provided in Algorithm~\ref{al:miy_sim}, with a straightforward
simulation of all components indices in parallel in Step~1 and a simulation of the parameters of the mixture exploiting
the conjugate nature of the prior against the complete-data  likelihood function in Step 2.
Some implementations of this algorithm for various distributions from the exponential family
can be found in the reference book of \cite{fruhwirth:2006}.\index{Gibbs sampling|)}

\paragraph*{Illustration}

As an illustration, consider the setting of a univariate Gaussian mixture with two components with equal
and known variance $\sigma^2$ and fixed weights  $(\eta,1-\eta)$:
\begin{equation}\label{eq:gimi9}
\eta\,\mathcal{N}(\mu_1,\sigma^2) + (1-\eta)\,\mathcal{N}(\mu_2,\sigma^2) \,.
\end{equation}
The only parameters of the model are thus $\mu_1$ and $\mu_2$.
We assume in addition a Normal $\mathcal{N}(0,10\sigma^2)$ prior distribution on both means $\mu_1$ and $\mu_2$.
Generating directly i.i.d. samples of $(\mu_1,\mu_2)$'s distributed according to  the posterior distribution associated with an
observed sample $\mathbf{y} = (y_1,\ldots,y_n)$ from (\ref{eq:gimi9})
quickly become impossible, as discussed for instance in \cite{Diebolt:Robert:1994} and \cite{Celeux:Hurn:Robert:2000},
because of a combinatoric explosion in the number of calculations, which grow as $\Omath(2^n)$. The posterior is
indeed solely interpretable as a well-defined mixture of standard distributions that involves that number of components.

\begin{figure}[t!]
\includegraphics[height=.9\textwidth,width=5cm,angle=270]{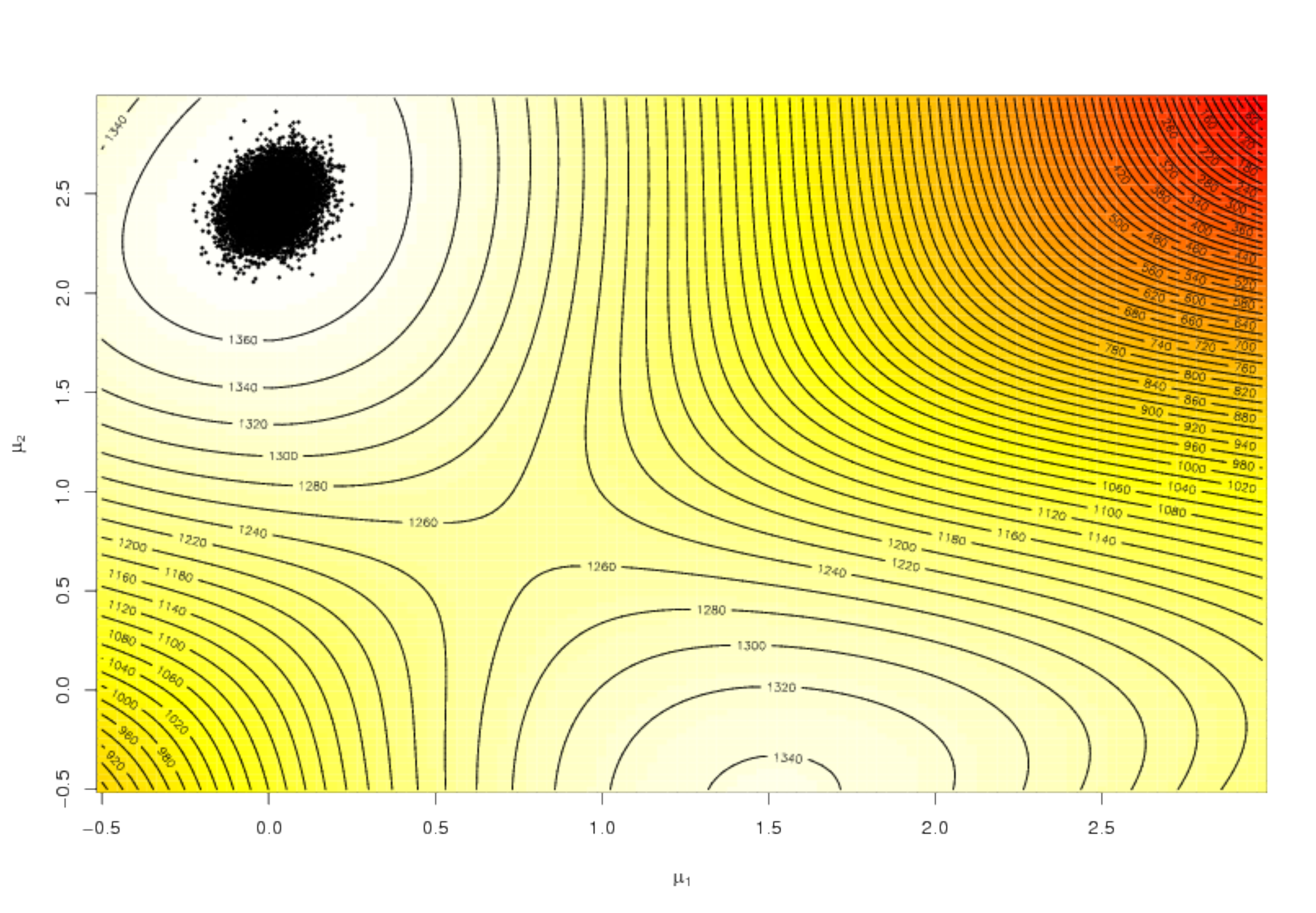}
\caption[Gibbs output for mixture posterior]{\label{fig:gib+mix1}
Gibbs sample of $5,000$ points for the mixture posterior against the posterior
surface. The level sets in the image are expressed in a logarithmic scale.
{\em (Reproduced with permission from \citealp{robert:casella:2004}.)}}
\end{figure}

\begin{figure}[t!]
\includegraphics[height=.9\textwidth,width=5cm,angle=270]{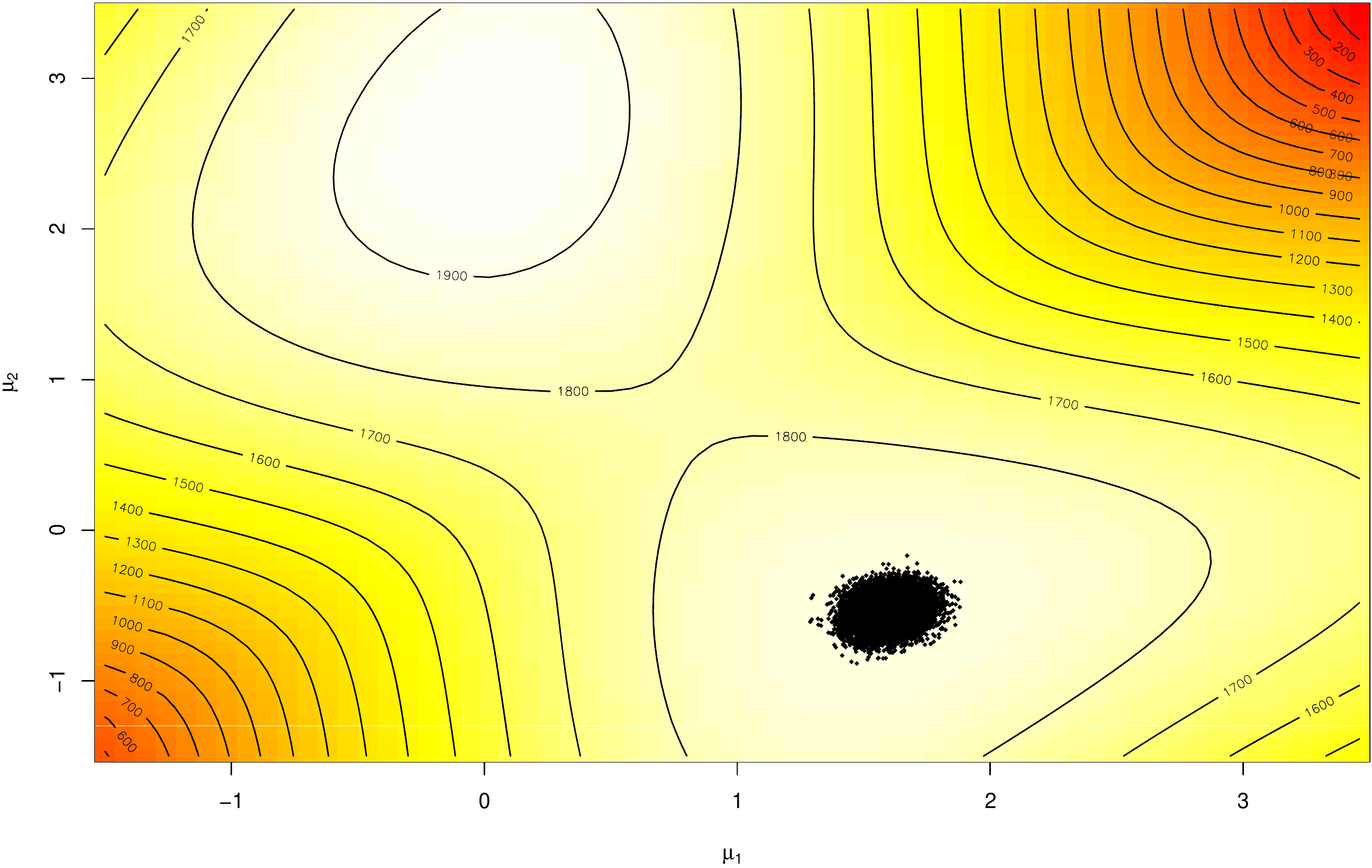}
\vskip -.3truecm
\caption[Gibbs sampler stuck at a wrong mode]{Gibbs sample for the two mean mixture
model, when initialised close to the second and lower mode, for true values
$\mu_1=0$ and $\mu_2=2.5$, over the log-likelihood surface.
{\em (Reproduced with permission from \citealp{robert:casella:2004}.)}}
\label{fig:wronmix}
\end{figure}

As explained above, a natural completion of $(\mu_1,\mu_2)$
is to introduce the (unobserved) component indicators $z_i$ of the observations $y_i$, in a similar way as for the EM algorithm,\index{EM algorithm!and Gibbs sampler} namely,
$$
\Prob (z_i=1) = 1- \Prob (z_i=2) = \eta \qquad\mbox{and}\qquad y_i|z_i=g\sim\mathcal{N}(\mu_g,\sigma^2) \,.
$$
The completed distribution with $\zm=(z_1,\ldots,z_n)$ is thus
\begin{align*}
p (\mu_1,\mu_2,\mathbf{z}|\mathbf{y}) &\propto \exp\{-(\mu_1^2+\mu_2^2)/20 \sigma^2 \}\,
\prod_{i;z_i=1} \eta \exp\{-(y_i-\mu_1)^2/2\sigma^2 \}\,\times\\
&\prod_{i;z_i=2} (1-\eta) \exp\{-(y_i-\mu_2)^2/2\sigma^2 \}\,.
\end{align*}
Since $\mu_1$ and $\mu_2$ are independent, given $(\mathbf{z},\mathbf{y})$,  the
conditional distributions are $(g=1,2)$:
$$
\mu_g|\ym \sim \mathcal{N}\left( \sum_{i;z_i=g} y_i / \left( 0.1 + n_g\right),
        \sigma^2 / \left( 0.1 + n_g\right) \right)\,,
$$
where $n_g$ denotes the number of $z_i$'s equal to $g$ and  0.1=1/10 represent the prior precision.
Similarly, the conditional distribution of $\mathbf{z}$ given
$(\mu_1,\mu_2)$ is a product of binomials, with
\begin{eqnarray*}
\lefteqn{\Prob(z_i=1|y_i,\mu_1,\mu_2)} \\
&=& \frac{\eta \exp\{-(y_i-\mu_1)^2/2\sigma^2 \} }{ \eta \exp\{-(y_i-\mu_1)^2/2\sigma^2 \} + (1-\eta)
\exp\{-(y_i-\mu_2)^2/2\sigma^2 \} }\,.
\end{eqnarray*}
Figure \ref{fig:gib+mix1} illustrates the behavior of the Gibbs sampler in that
setting, with a simulated data set of $5,000$ points from the $0.7 \mathcal{N}(0,1)
+0.3 \mathcal{N}(2.5,1)$ distribution. The representation of the MCMC sample after
$15,000$ iterations is quite in agreement with the posterior surface, represented via
a grid on the $(\mu_1,\mu_2)$ space and some contours; while it may appear to be too
concentrated around one mode, the second mode represented on this graph is much lower
since there is a difference of at least $50$ in log-posterior units.
However, the Gibbs sampler may also fail to converge\index{Gibbs sampling!convergence}, as described in
\cite{Diebolt:Robert:1994} and illustrated in Figure \ref{fig:wronmix}. When
initialised at the secondary mode\index{multimodality}\index{mixtures!modes} of the likelihood, the magnitude of the moves around
this mode may be too limited to allow for exploration of further modes
(in a realistic number of iterations).

\paragraph*{Label switching}

Unsurprisingly,\index{label switching} given the strong entropy of the local
modes demonstrated in the illustrative example, Gibbs sampling performs very
poorly in terms of label switching (already discussed in
Chapter~1) in that it rarely switches between equivalent modes of
the posterior distribution. One (if not the only)  reason for this behaviour is
that, due to the allocation of the observations to the various components,
i.e., by completing the unknown parameters with the unknown (and possibly
artificial) latent variables, the  Gibbs sampler faces enormous difficulties in
switching between equivalent modes, because this amounts to changing the values
of most latent variables to a permuted version all at once. It can actually be
shown that the probability of seeing a switch goes down to zero as the sample
size $n$ goes to infinity.  This difficulty is increasing with dimension, in
the sense that the likelihood function increasingly peaks with the dimension.

Some \citep[as, e.g.,][]{geweke:2007} would argue the Gibbs sampler does very well by naturally selecting a mode and sticking to it. This certainly
favours estimating the parameters of the different components. However, there is no guarantee that the Gibbs sampler will
remain in the same mode over iterations.

The problem can alleviated to a certain degree by enhancing  the Gibbs sampler (or any other simulation based technique) to switch between equivalent modes. A simple, but efficient, solution to obtain a sampler that explores all symmetric modes of  the posterior distribution is to enforce balanced label switching by concluding each MCMC draw by a random permutation of the labels. Let $\mathfrak{S}(G)$ denote  the set of the $G!$ permutations of $\{1,\ldots,G\}$. Assume that $ (\eta_1,\ldots,\eta_G)$ follows the symmetric
 Dirichlet distribution  $\Dirinv{G}{e_0}$
 (which  corresponds to the $\Dir(e_0, \ldots, e_0)$ distribution and
 is invariant to permuting the labels by definition) and assume that also  the prior on $\theta_g$ is invariant in this respect, i.e.  $p(\theta_{\perm(1) }, \ldots, \theta_{\perm(G) })=p(\theta_1, \ldots, \theta_G)$ for all permutations $\perm \in \mathfrak{S}(G)$. Then, for any given posterior draw $\theta=(\eta_1,\ldots,\eta_G, \theta_1, \ldots, \theta_G)$,   jumping between the equivalent modes of the posterior distribution  can be achieved by
 defining the permuted draw   $\perm(\theta):=(\eta_{\perm(1)},\ldots,\eta_{\perm(G)}, \theta_{\perm(1)}, \ldots, \theta_{\perm(G)})$ for some permutation $\perm \in \mathfrak{S}(G)$.

 This idea underlies the random permutation sampler introduced by  \citet{fruh:2001}, where each  of the $T$ sweeps of Algorithm~\ref{al:miy_sim} is concluded by such a permutation of the labels, based on randomly selecting one of the $G!$ permutations  $\perm \in \mathfrak{S}(G)$, see also \citet[Section~3.5.6]{fruhwirth:2006}.
 Admittedly, this  method  works well only, if  $G ! \ll T$, as the expected number of draws from each modal region  is equal to $T/G !$.
  \citet{geweke:2007} suggests to consider {\em all} of the  $G!$ permutations in $\mathfrak{S}(G)$ for each of the $T$ posterior draw, leading to a completely balanced sample of $T \times G!$ draws, however the resulting sample size can be enormous, if $G$ is large.

As argued in Chapter~1, the difficulty in exploring all modes of the posterior distribution in the parameter space is not a primary concern provided the space of mixture distributions (that are impervious to label-switching) is correctly explored. Since this is a space that is much more complex than the Euclidean parameter space, checking proper convergence is an issue. Once again, this difficulty is more acute in larger dimensions and it is compounded by the fact that secondary modes of the posterior become so \lq\lq sticky\rq\rq\  that a standard Gibbs sampler cannot escape their (fatal) attraction. It may therefore be reasonably argued, as in \cite{Celeux:Hurn:Robert:2000} that off-the-shelf Gibbs sampling does not necessarily work for mixture estimation, which would not be an issue in itself were alternative generic solutions readily available!

Label switching, however, still matters very much when considering the
statistical evidence\index{evidence|see{marginal likelihood}}\index{marginal likelihood} (or marginal likelihood)
associated\index{likelihood!marginal}\index{marginal likelihood}
with a particular mixture model, see also Chapter~7. This evidence\index{Chib's evidence approximation}
can be easily derived from Bayes' theorem by Chib's (\citeyear{chib:1995})
method, which can also be reinterpreted as a Dickey-Savage representation\index{Dickey-Savage representation}
\citep{dickey:1968}, except that the Rao-Blackwell representation of the
posterior distribution of the parameters is highly dependent on a proper mixing
over the modes. As detailed in \cite{neal:1999} and expanded in
\cite{fruhwirth:2004}, a perfect symmetry must be imposed on the posterior
sample for the method to be numerically correct. In the most usual setting when
this perfect symmetry fails, it must be imposed in the estimate, as proposed in
\cite{berkhof:mechelen:gelman:2003} and \cite{lee:robert:2015}, see also Chapter~7 for more details.

\subsection{Metropolis--Hastings schemes}

As detailed at the beginning of this chapter, computing the likelihood function at a given parameter value $\theta$ is
not a computational challenge provided (i) the component densities are available in closed-form\footnote{Take, e.g., a mixture
of $\alpha$-stable distributions as a counter-example.} and (ii) the sample size remains manageable, e.g., fits within a
single computer memory. This property means that an arbitrary proposal can be used to devise a Metropolis--Hastings
algorithm \citep{robert:casella:2004,lee:mari:meng:robe:2008} associated with a mixture model, from a simple random walk
to more elaborate solutions like Langevin and Hamiltonian Monte Carlo.\index{Metropolis-Hastings algorithm}

The difficulty with this Metropolis--Hastings approach is to figure out an efficient way of implementing the simulating
principle, which does not provide guidance on the choice of the proposal distribution. Parameters are set in
different spaces, from the $G$-dimensional simplex of $\mathbb{R}^G$ to real vector spaces.
A random walk is thus delicate to
calibrate in such a context and it is often preferable to settle for a Gibbs sampler that updates one group of
parameters at a time, for instance the weights, then the variances, then the means in the case of a location-scale
mixture. This solution however shares some negative features with the original
Gibbs sampler in the sense that it may prove a hardship to explore the entire parameter
space (even without mentioning the label switching problem). Still, the
implementation of the unidimensional location-parameterisation of
\cite{kamary:lee:robert:2015} 
relies on this
block-wise version and manages to handle a reasonably large number of
components, if not a larger number of dimensions. Indeed, when the dimension $d$ of the observables increases, manipulating
$d\times d$ matrices gets particularly cumbersome and we know of no generic solution to devise an automatic random walk Metropolis--Hastings approach in this case.

Among the strategies proposed to increase the efficiency of a
Metropolis--Hastings algorithm in dimension one, let us single out the
following ones:
\begin{enumerate}
\item Deriving independent proposals based on sequential schemes starting for
instance from maximum likelihood and other classical estimates, since those are
usually fast to derive, and followed by mixture proposals based on subsamples,
at least in low dimensional models;
\item An overparameterisation of the weights $\eta_g$ defined as\index{overparameterisation}
$\eta_g=\alpha_g/\sum_j \alpha_j$, with a natural extension of the Dirichlet prior
$\mathcal{D}(e_0,\ldots,e_0)$ into a product of $d$ Gamma priors  on the $\alpha_g$'s, where $\alpha_g \sim \mathcal{G}(e_0,1)$.
 This representation avoids restricted simulations
over the simplex of $\mathbb{R}^G$ and adding an extra parameter means the
corresponding Markov chain mixes better;
\item The inclusion of the original Gibbs steps between random exploration proposals, in order to thoroughly survey the
neighbourhood of the current value. Gibbs sampling indeed shares to some extent the same property as the EM algorithm to
shoot rapidly towards a local mode when started at random. In this respect, it eliminates the random walk poor behaviour
of a standard Metropolis--Hastings algorithm. Cumulating this efficient if myopic exploration of the Gibbs sampler
with the potential for mode jumping associated with the Metropolis--Hastings algorithm may produce the best of two
worlds, in the sense that mixture proposals often achieve higher performances than both of their components \citep{tierney:1994};
\item Other mixings of different types of proposals, using for instance
reparameterisation, overparameterisation, or underparameterisation. One
important example is given by \cite{rousseau:mengersen:2011}. In this paper,
already discussed in Chapters~1 and 4, the authors consider mixtures with
\lq\lq too many\rq\rq\ components and demonstrate that a Bayesian analysis of such
overfitted models manages to eliminate the superfluous components. While this
is an asymptotic result and while it imposes some constraints on the choice of
the prior distribution on the weights, it nonetheless produces a theoretical
warranty that working with more components than needed is not ultimately
damaging to inference, while allowing in most situations for a better mixing
behaviour of the MCMC sampler;\index{overfitted mixtures}
\item Deriving nonparametric estimates based for instance on Dirichlet process mixtures returns a form of clustering that can be exploited to build component-wise proposal;
\item Approximate solutions like nested sampling (see below), variational
Bayes \citep{zobay:2014}, or Expectation--Propagation (EP, \citealp{minka:2001,titterington:2011}) may lead to independent proposals that contribute to a larger
degree of mobility over the posterior surface;
\item Further sequential, tempering, and reversible jump solutions as discussed below.
\end{enumerate}
At this stage, while the above has exhibited a medley of potential fixes to the Metropolis--Hasting woes, it remains urgent to
warn the reader that no generic implementation is to be found so far (in the sense of a generic software able to handle
a wide enough array of cases). The calibration stage of those solutions remains a challenging issue that hinders and in some cases prevents an MCMC resolution of the computational problem. This is almost invariably the case when the dimension of the model gets into double digits,  see Section~\ref{secprihigh}.

\subsection{Reversible jump MCMC}

It may seem inappropriate to include a section or even a paragraph on\index{reversible jump MCMC}\index{RJMCMC}
reversible jump MCMC \citep{green:1995} in this chapter since we are not
directly concerned with estimating the number of components, however, this
approach to variable or unknown $G$ environments is sometimes advanced as a
possible mean to explore better the parameter space by creating passageways
through spaces of larger (and possibly smaller) dimensions and numbers of
component. As discussed in \cite{chopin:robert:2010}, this strategy is similar
to bridge sampling. Once again, calibration of the method remains a major endeavour \citep{richardson:green:1997},
especially in multidimensional settings, and we thus abstain from describing this
solution any further.

\subsection{Sequential Monte Carlo}

Sequential Monte Carlo methods (see \citealp{delmoral:doucet:jasra:2006})\index{sequential Monte Carlo}
approach posterior simulation by a sequence of approximations, each
both closer to the distribution of interest and borrowing strength from the
previous approximation. They therefore apply even in settings where the data is\index{particle filtering}
static and entirely available from the start. They also go under the names of particle systems and particle filters.

Without getting into a full description of the way particle systems operate, let us
recall here that this is a particular type of iterated importance sampling
where, at each iteration $t$ of the procedure, a weighted sample
$(\theta_{1t},\ldots,\theta_{Nt})$ of size $N$ is produced, with weights $\omega_{it}$
targeting a distribution $\pi_t$.
The temporal and temporary target $\pi_t$
may well be supported by another space than the support of the original target
$\pi$. For instance, in \cite{everitt:etal:2016}, the $\pi_t$'s are the
posterior distributions of mixtures with a lesser number of components, while
in \cite{chopin:2002} they are posterior distributions of mixtures with a lesser
number of observations.\footnote{One could equally conceive the sequence of targets as
being a sequence of posterior distributions of mixtures with a lesser number of dimensions or with lesser correlations structure, for instance borrowing from variational Bayes.}
The sample or {\em particle system} at time $t$ is instrumental in designing
the importance proposal for iteration $t+1$, using for instance MCMC-like
proposals for simulating new values. When the number of iterations is large
enough and the temporary targets $\pi_t$ are too different, the importance weights
necessarily deteriorate down to zero (by basic martingale theory) and particle
systems include optional resampling steps to select particles at random based
on the largest weights, along with a linear increase in the number of particles
as $t$ grows.  The construction of the sequence of temporary targets $\pi_t$ is
open and intended to facilitate the exploration of intricate and highly
variable distributions, although its calibration is delicate and may jeopardise
convergence.

In the particular setting of Bayesian inference, and in the case of mixtures,
a natural sequence can be associated with subsample posteriors, namely
posteriors constructed with only a fraction of the original data, as proposed for instance in
\cite{chopin:2002,chopin:2003}. The starting
target $\pi_0$ may for instance correspond to the true prior or to a posterior
with a minimal sample size \citep{robert:2007}. A more generic solution is to
replace the likelihood with small powers of the likelihood in order to flatten
out the posterior and hence facilitate the exploration of the parameter space.
A common version of the proposal is then to use a random walk, which can be
calibrated in terms of the parameterisation of choice and of the scale based on
the previous particle system. The rate of increase of the powers of the likelihood can also be
calibrated in terms of the degeneracy rate in the importance weights. This
setting is studied by \cite{everitt:etal:2016} who point out the many
connections with reversible jump MCMC.

\subsection{Nested sampling}     \label{sec5nest}

While nested sampling is a late-comer to the analysis of mixtures \citep{skilling:2004}, one of the first examples
of using this evidence estimation technique is a mixture example. We cannot  recall the basics and background
of this technique here and simply remind the reader that it consists of simulating a sequence of particles over
subsets of the form\index{nested sampling}
$$
\left\{\theta; \lik (\theta|\by)\ge \alpha\right\}
$$
where $\lik$ is the likelihood function and $\alpha$ a bound updated at each
iteration of the algorithm, by finding the lowest likelihood in the current
sample and replacing the argument with a new value with a higher likelihood.
For a mixture model this approach offers the advantage of using solely the numerical value of the likelihood at a given
parameter value, rather than exploiting more advanced features of this function. The resulting drawback is that the
method is myopic, resorting to the prior or other local approximations for proposals. As the number of components, hence
the dimension of the parameter, increases, it becomes more and more difficult to find moves that lead to higher values
of the likelihood. In addition, the multimodality of the target distribution implies that there are more and more parts
of the space that are not explored by the method
\citep{chopin:robert:2010,marin:robert:2010}. While dimension (of the mixture
model as well as of the parameter space) is certainly an issue and presumably a
curse \citep{buchner:2014} the Multinest version of the algorithm manages
dimensions up to 20 \citep{feroz:etal:2013}, which remains a small number when
considering multivariate mixtures.\footnote{A bivariate Gaussian mixture with
four components involves more than 20 parameters.}

\section{Bayesian Inference in the Model-based Clustering Context} \label{sec:bimbclust}

In addition to being a probabilistic model {\em per se}, finite mixture models
provide a well-known probabilistic approach to clustering. In the model-based
clustering setting each cluster is associated with a mixture component. Usually
clustering is relevant and useful for data analysis  when the number of
observations is large, involving say several hundreds of observations, and so is
the number of variables, with say several dozens of variables. Moreover,
choosing the unknown number $G$ of mixture components corresponding to the data
clusters is a sensitive and critical task in this settings. Thus efficient
Bayesian inference for model-based clustering requires MCMC algorithms working
well and automatically in large dimensions with potentially numerous
observations, which themselves  requires smart strategies to derive a relevant number of
clusters, see Chapter~7 for a detailed discussion.\index{Bayesian!cluster analysis}

\cite{MWFSG16} devoted considerable efforts to assess relevant Bayesian procedures in a
model-based clustering context and we refer the reader to this paper for detailed coverage.
In this section, we only summarise their inference strategy,
which consists primarily of choosing relevant prior distributions. As an illustration, their
approach is implemented in the following section in a realistic if specific case with
fifty variables and a relatively large number of observations.

We recall that the goal of the approach is to cluster $n$ individuals made of $d$ quantitative variables.
In \cite{MWFSG16}, each cluster is associated with a multivariate Gaussian
distribution, resulting formally in a multivariate Gaussian mixture sample with $G$ components
$$
\sum_{g=1}^G \eta_g\mathcal{N}_d\left(\mu_g,\Sigma_g\right)\,,
$$
where $G$ is typically unknown.\index{prior!selection|(}

\paragraph{Choosing priors for the mixing proportions}
\cite{MWFSG16}'s strategy consists of starting the analysis with
an overfitted mixture, that is, a mixture with a number of components $G$ most likely
beyond the supposed (if unknown) number of relevant clusters. Assuming a
symmetric Dirichlet prior $\Dirinv{G}{e_0}$  on the mixing
proportions, they argue, based upon asymptotic results established in
\cite{rousseau:mengersen:2011} and as observed in \cite{fruh11}, that if
$e_0<r/2$, $r$ being the dimension of the component-specific parameter
$\theta_g$, then the posterior expectation of the mixing proportions converges
to zero for superfluous components. But if $e_0>r/2$ then the posterior density
handles overfitting by defining at least two identical components, each with
non-negligible weights. Thus \cite{MWFSG16} favor small values of $e_0$ to
allow emptying of superfluous components. More precisely, they consider a hierarchical prior distribution, namely a
symmetric Dirichlet prior $\Dirinv{G}{e_0}$ on the component weight distribution  $(\eta_1, \ldots, \eta_G)$ and
 a Gamma hyperprior for $e_0$:
\begin{eqnarray*}
\eta_1, \ldots, \eta_G |e_0 \sim \Dirinv{G}{e_0}, \qquad   e_0 \sim   \Gammad (a, a G).
\end{eqnarray*}
Based on numerical experiments on simulated data, they recommend setting $a=10$.
An alternative choice  
is fixing the hyperparameter $e_0$ to a given small value.

\paragraph{Choosing the priors for the component means and covariance matrices}
Following \cite{fruh11}, \cite{MWFSG16} recommend putting  a shrinkage prior, namely the Normal Gamma prior on component means.
This prior is designed to handle high-dimensional mixtures, where not all variables contribute to the clustering structure, but a number of irrelevant variables is expected to be present without knowing a priori which variables this could be. For any such variable $y_{il}$ in dimension $l$, the components means $\mu_{l1}, \ldots, \mu_{lG}$ are pulled toward a common  value $b_{0l}$, due to a  local, dimension-specific shrinkage  parameter $\lambda_l$.
More specifically,  the following hierarchical prior  based on a   multivariate Normal distribution for the component means $\mu_g$ is chosen:
\begin{eqnarray*}\label{eq:hier}
& \displaystyle \mu_g \sim \Normal_d(b_0, B_0), \quad g=1, \ldots, G, & \\
& \displaystyle b_0 \sim \Normal_d(m_0, M_0) ,  \quad B_0=\Lambda R_0 \Lambda,  \quad R_0 =\diag (R_1^2, \ldots, R_d^2), \\
& \displaystyle \Lambda =\diag  (\sqrt{\lambda_1}, \ldots, \sqrt{\lambda_d}), \quad \lambda_l\sim \Gammad(\nu_1, \nu_2), \quad l=1, \ldots, d,
\end{eqnarray*}
where $R_l$ is the range of $y_{il}$ (across $i=1,\ldots, n$).
\cite{MWFSG16} suggest setting the hyperparameters $\nu_1$ and $\nu_2$  to
$0.5$ to allow for a sufficient shrinkage of the prior variance of the
component means. For $b_0$, they specify an improper and empirical prior where
$m_0=\text{median}(y)$ and $M_0^{-1}=0$.

	\begin{figure}[t!]
	\centering
	\includegraphics[width=0.32\textwidth]{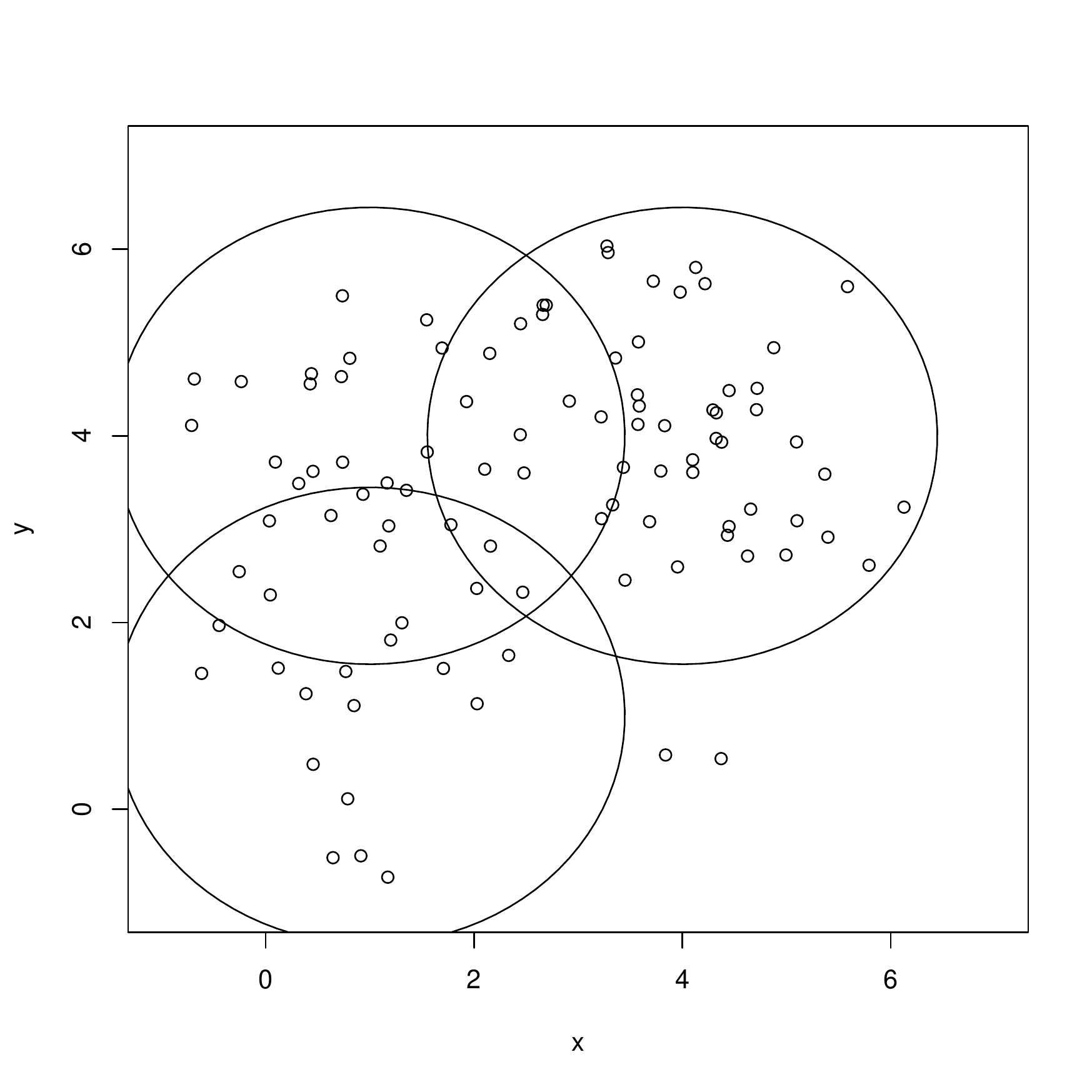}
	\includegraphics[width=0.32\textwidth]{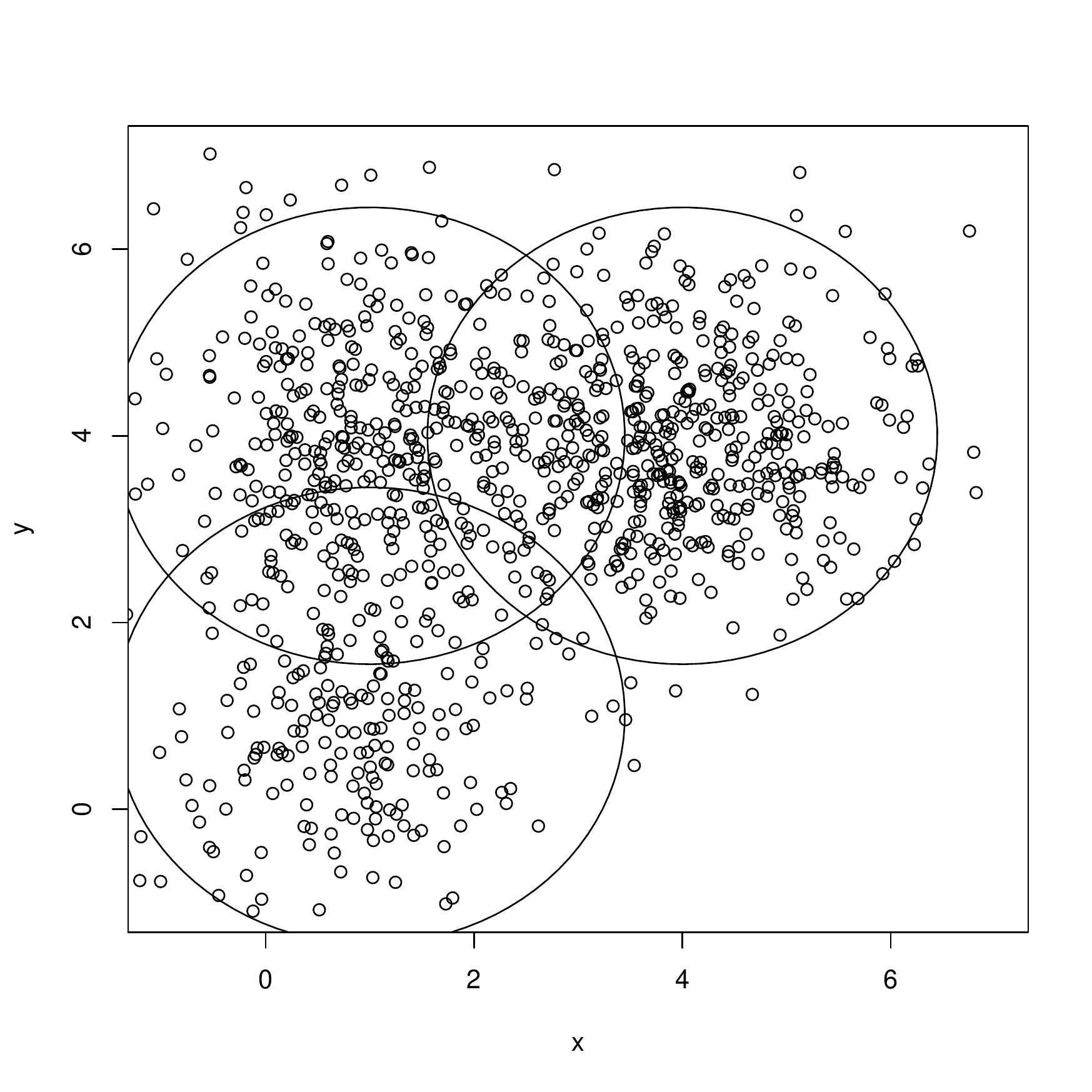}
	\includegraphics[width=0.32\textwidth]{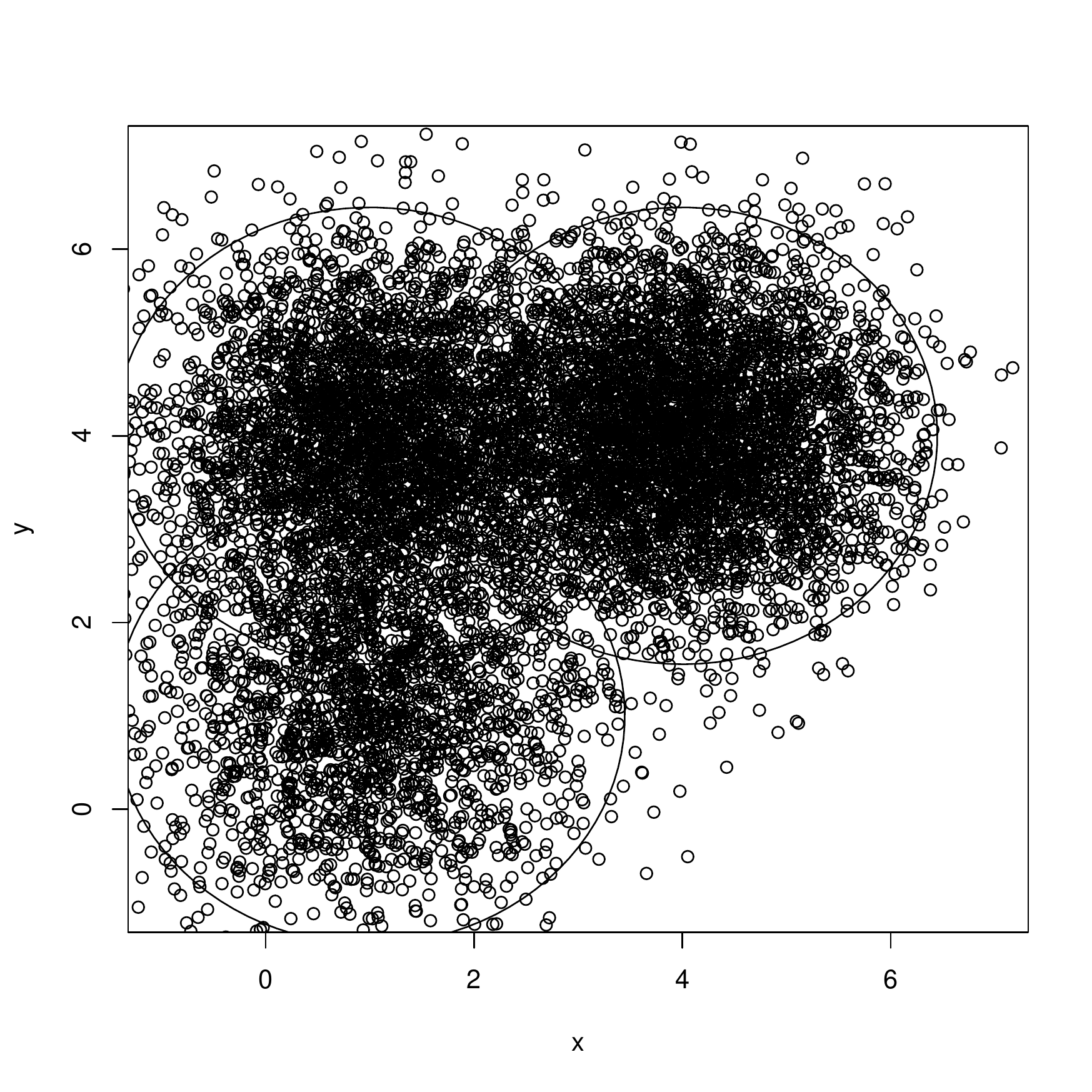}
	\vskip -.3truecm
	\caption{Scatterplots  of  simulated data sets  (first two dimensions),  with $n=100,1000,10000$ respectively and circles denoting the 95\% probability regions for each component of mixture (\ref{eq:1}).}
	\label{plot:data}
\end{figure}

A standard conjugate hierarchical prior is considered on the component covariance matrices $\Sigma_g$:\footnote{As opposed to the other chapters,
we use here the same parametrizations of the Wishart and Inverse Wishart distribution as employed in \citet{fruhwirth:2006}:  $\Sigma_g^{-1}\sim \Wishart (c_0,C_0)$ iff $\Sigma_g \sim \Wishartinv (c_0,C_0)$, with $\Ew(\Sigma_g^{-1})=c_0 C_0^{-1}$ and $\Ew(\Sigma_g)=C_0/(c_0-(d+1)/2)$. In this parametrization,  the standard Gamma and Inverse Gamma distribution are recovered when $d=1$.}
\begin{align}\label{eq:IW}
\begin{aligned}
\Sigma_g^{-1}&\sim \Wishart(c_0, C_0), \quad C_0\sim \Wishart(g_0, G_0), \\
c_0=2.5+\frac{d-1}{2}, \quad g_0&=0.5+\frac{d-1}{2}, \quad G_0=\frac{100 g_0}{c_0}\diag (1/R_1^2, \ldots, 1/R_d^2).
\end{aligned}
\end{align}
Under such prior distributions, an MCMC sampler is detailed in Appendix~1
of \cite{MWFSG16}. 
  The point process representation of
the MCMC draws introduced in \cite{fruhwirth:2006} and recalled in Chapter~1 is exploited to study the
posterior distribution of the component-specific parameters, regardless of
potential label switching. This is achieved through a $G$-centroids cluster
analysis based on Mahalanobis' distance as detailed in Appendix~2 of \cite{MWFSG16}.

Based on the simulation experiments conducted in Section~\ref{chaptersim}, we conclude that  prior (\ref{eq:IW})
is problematic for high-dimensional mixtures with large values of $d$. In Section~\ref{secprihigh}, a modification of
prior (\ref{eq:IW}) is presented that works well  also for very high-dimensional mixtures.

A final remark is that  in a model-based clustering context  the
means of the various clusters are expected to be distinct. It is thus advisable to initialise
the  MCMC algorithm with a $k$-means algorithm in $\mathbb{R}^d$ in order to
make sure that the Markov chain mixes properly and to avoid being stuck in a slow
convergence area. Obviously, in large or even moderate dimensions, there is a
clear need for performing several runs of the MCMC algorithm from different
random positions to ensure that the MCMC algorithm has reached its stationary
distribution.\index{prior!selection|)}

\section{Simulation Studies}   \label{chaptersim}

In this section, we study the specific case of a multivariate Gaussian mixture model with three components
in dimension $d=50$. We simulate observations  $\by=(y_1,\ldots,y_n)$   from $y \in\mathbb{R}^d$ such that
\begin{equation}\label{eq:1}
y \sim \sum_{g=1}^3 \eta_g \mathcal{N}_d(\mu_g, \Sigma_g)\,
\end{equation}
\noindent with $(\eta_1, \eta_2, \eta_3)=(0.35,0.2,0.45)$,
$\mu_1=(1, 4,\ldots,1,4) ^\top $, $\mu_2=(1,1,\ldots,1) ^\top $, $\mu_3=(4,4,\ldots,4) ^\top $
and $\Sigma_g=\tau I_d$ ($g=1,2,3$). The parameter $\tau$ is chosen in the
simulations to calibrate the overlap between the components of the mixture.
Examples of data sets  are shown in Figure \ref{plot:data} for varying $n$.

The simulated distribution is homoscedastic and the true covariance matrices are isotropic.
However, we refrained from including this information in the prior
distributions to check  the performance of the proposed samplers.
A direct implication of this omission is that the three covariance matrices
involve $3\left(\frac{d\times(d-1)}{2}+d\right)=3,825$ parameters instead of
a single parameter. Taking furthermore into account the three component means  and
the mixture weights, the dimension of the parameter space increases to $3,977$.

It would be illuminating  to fit a mixture model to the simulated data sets
using the various approaches outlined in Section~\ref{section2Alg}  and  to compare the computational performance of the different
algorithms. However,  our first observation is that the Gibbs sampler appears
to be the sole manageable answer to generate parameter samples from the
posterior distribution in such a high-dimensional experiment.  Alternative
methods like Metropolis-Hasting schemes and sequential Monte Carlo samplers
turned out to be extremely delicate to calibrate and, despite some significant
investment in the experiment, we failed to find a satisfying calibration of the
corresponding tuning parameters in order to recover the true parameters used
for simulating the data.  For this reason,  investigation of the sampling
methods is limited to Gibbs sampling for the remainder of this chapter.

A second interesting finding is that the choice of  the scale parameter $\tau$
in (\ref{eq:1})  played a major role in the ability of the Gibbs sampler to recover
the true parameters. In particular for simulations associated with $\tau\geq 5$
and a sample size between $n=100$ and $n=500$, meaning a significant overlap,
the choice of the hyperparameters of the inverse Wishart prior  (\ref{eq:IW})
on the covariance matrices has a considerable impact on the shape and bulk of the posterior distribution.

\begin{table}[t!]
\caption{Estimated number of components $\hat{G}$ for a  data set of $n$ observations drawn from the three component mixture (\ref{eq:1}), averaged number across $M=30$ independent replicates of Gibbs sampling  with $T$ iterations}
\label{tab:kaniav}
\begin{center}
\begin{tabular}{cccc}
\hline
$T$   & $n $ &  $\min(\hat{G})$  & $\max(\hat{G})$  \\
\hline
$10^3$   & 100    & 10 & 10 \\
$10^4$   & 100    & 10 & 10 \\
$10^3$   & 500    & 10 & 10 \\
$10^4$   & 500    & 10 & 10 \\
$10^4$   & 1,000  & 10 & 10 \\
$10^3$   & 10,000 & 3  & 3  \\
\hline
\end{tabular}
\end{center}
\end{table}

Using the Matlab library {\sf bayesf}\index{bayesf@{\sf bayesf}} associated with the book
of \cite{fruhwirth:2006}, we observe poor performances of default choices,
namely,  either the conditionally conjugate prior proposed by
\cite{bensmail:celeux:raftery:robert:1997} or the hierarchical independence
prior introduced by \cite{richardson:green:1997} for univariate Gaussian
mixture and extended by \cite{stephens:1997} to the multivariate case, see
Section 6.3.2 of \cite{fruhwirth:2006} for details on these priors. For such priors, we indeed found that inference
led to a posterior distribution that is concentrated quite far from the
likelihood.  The reason for the discrepancy is that the posterior mass accumulates in regions of the parameter space
where some mixture weights are close to zero.  Thanks to the assistance of Sylvia
Fr\"uhwirth-Schnatter, we managed to fix the issue for the hierarchical
independence prior. Note that this prior distribution happens to be quite similar to the one
introduced by \cite{MWFSG16} and described in the previous section. We found through trial-and-error that
we had to drastically increase the value of $c_0$ in the inverse Wishart prior (\ref{eq:IW})
to, e.g., $c_0=50+\frac{d-1}{2}$, instead of the default choice
$c_0=2.5+\frac{d-1}{2}$. The central conclusion of this experiment remains the fact that
in high-dimensional settings prior distributions are immensely delicate to calibrate. A new fully automatic way to
choose the hyperparameters of  (\ref{eq:IW}) in a high-dimensional setting will be discussed in Section~\ref{secprihigh}.

On the other hand,  experiments for  $\tau\leq 5$   indicate  that  concentration difficulties tend to vanish. In the following experiment, we chose the value $\tau=1$ which produced well-separated clusters for  illustration.

\subsection{Known number of components} \label{sec:known}

Recall that the proposal of \cite{MWFSG16} has been designed to estimate both the number of components and
the Gaussian expectations and covariances. However, the proposed methodology can easily be implemented
to estimate the  parameters values, while  the number of components is supposed to be known. The only modification is
to adjust the prior hyperparameters by  setting $G=3$ (which is the true value) and specify a fixed value $e_0=0.01$.\footnote{Although the later value actually assumes that the specified mixture model is overfitting (which is  not the case for these investigations), this prior setting still worked well in our simulations. Obviously, this observation remains conditional on these simulations.}

In order to evaluate whether the Gibbs sampler is able to recover the true component parameters, we simulate two data sets from (\ref{eq:1}), with $n=100$ and $n=1000$ observations, respectively. For these two data sets, we repeat running the Gibbs sampler $M=10$ times with $T=10,0000$  iterations
after a burn-in period of $1,000$ iterations. The posterior estimates are computed after a reordering
process based on the $G$-centroids clustering of $T$ simulated component means in the point process representation.

In order to check the sensitivity of the Gibbs sampling on the choice of the initial values for the MCMC algorithm,
hence indicating potential issues with convergence, we compared three different methods of initialisation:
\begin{itemize}
\item[(a)] {\bf Initialisation 1} determines initial values by performing $k$-means clustering of the data;
\item[(b)] {\bf Initialisation 2} considers maximum likelihood estimates   of the component parameters  computed by the \texttt{Rmixmod, https://cran.r-project.org/web/packages/Rmixmod/} package as initial values;
\item[(c)] {\bf Initialisation 3} allocates a random value to each parameter, simulated from the corresponding prior distribution.
\end{itemize}
Both  for $n=100$ and $n=1,000$, the posterior estimates obtained by implementing the  Gibbs sampler  based
on the three different initialisation  methods explained above are all very similar, as shown by our experiments
where ten repeated calls to the Gibbs sampler showed no visible discrepancy between the three posterior estimates.
This observed stability of the resulting estimations in terms of
initial values is quite reassuring from the point of view of the convergence of
the Gibbs sampler. (We have however to acknowledge that it is restricted to a
specific mixture model.) Furthermore, the component-wise parameter estimates associated with all three
initialisation methods  are all close to the corresponding true values.




\subsection{Unknown number of components}  \label{secunkonw}

In this section, we consider the joint  estimating both the number of components and the parameters values for various  data sets simulated from (\ref{eq:1}) with different numbers of observations $n$.
We implement the same Gibbs sampler as used in Section  \ref{sec:known}, however with $e_0=0.0001$  and  the maximum number of components being equal to $G=10$.  According to \cite{MWFSG16}, this prior setting empties the redundant components and the unknown  number of components can be estimated by the most frequent number of non-empty clusters, see also Chapter~7.  For both data sets, we run $M=30$ independent Gibbs samplers  with $T=10,000$  iterations after a burn-in period of $1,000$, using an initial clustering of the data points into 10 groups obtained through $k$-means clustering.

 As shown in Table~\ref{tab:kaniav}, when $n=10^4$ is large, then the method of \cite{MWFSG16}
always manages to pinpoint the true value of the number of mixture components (which is equal to three) in all replicas of our Gibbs sampler.
However,  for smaller data sets with $n$ ranging from $100$ to $1,000$,
the estimation of $G$ produces an overfit  of the true value. Even when the number of iterations was increased to $T=10^4$,
  the superfluous components did not get emptied during MCMC sampling, mainly because  the Gibbs sampler got  stuck in the initial classification.
  This hints at a prior-likelihood conflict in this high-dimensional setting which cannot be overruled for small data sets.
Therefore,  the following section  proposes an idea how to enforce convergence of the Gibbs sampler for high-dimensional mixtures through the specification of a  \lq\lq suitable\rq\rq\  prior on the component covariance matrices.

\section{Gibbs sampling  for high-dimensional mixtures}  \label{secprihigh}

\begin{figure}[t!]
		\centering
			\includegraphics[width=0.38\textwidth]{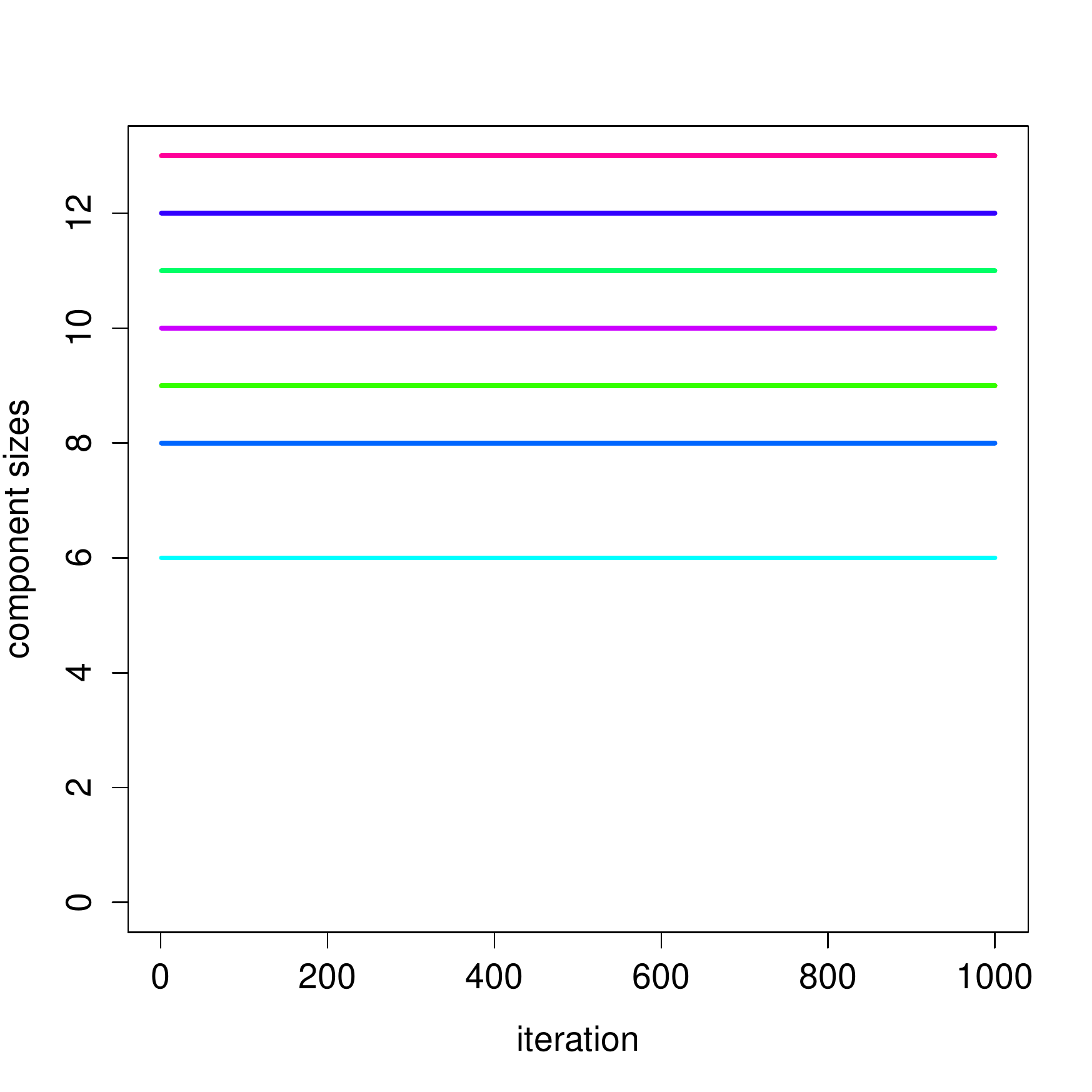}
			\includegraphics[width=0.38\textwidth]{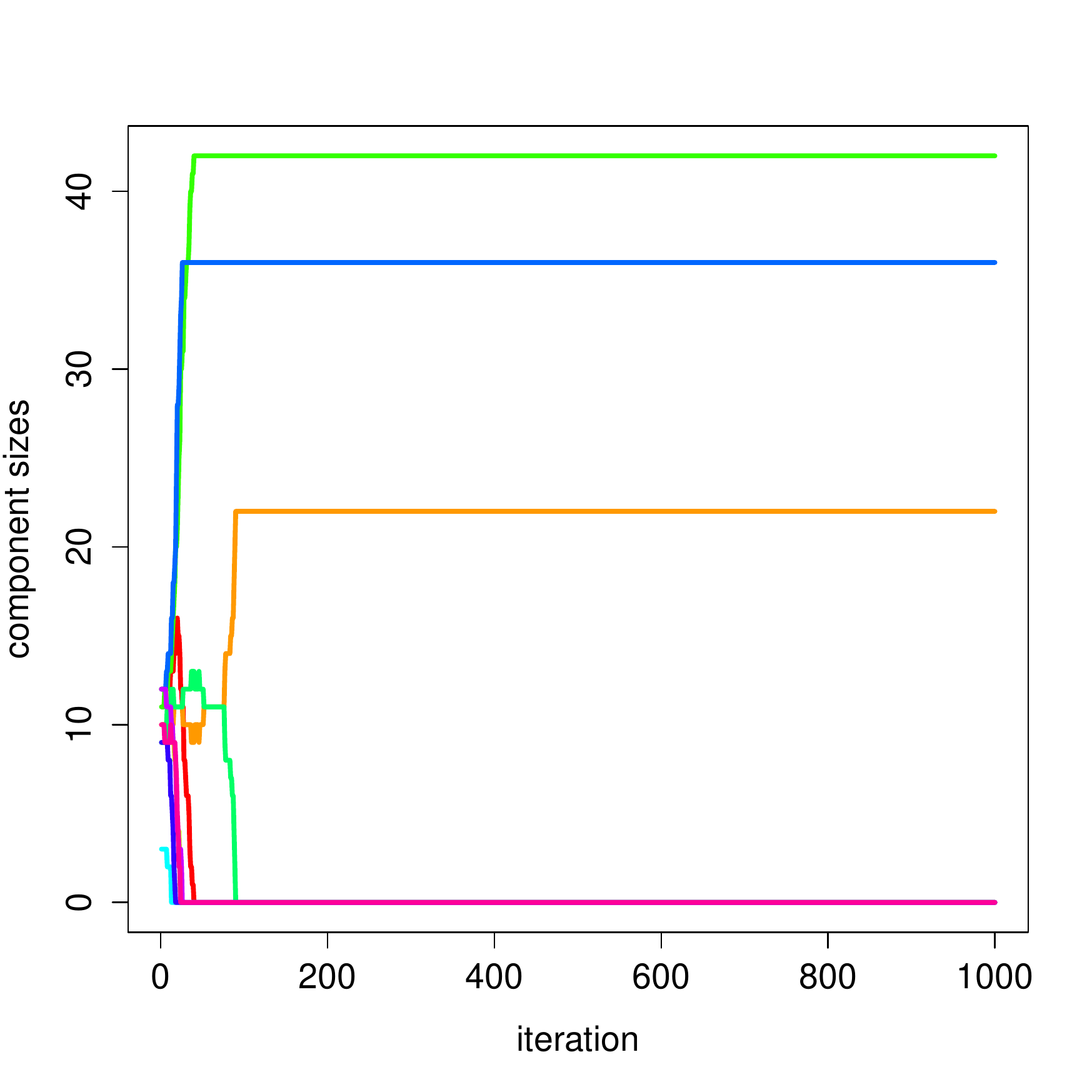}
	\vskip -.3truecm
		\caption{Trace plot of the number of observations assigned to the 10 components during MCMC sampling (each line corresponds to a component), under the prior used in \cite{MWFSG16}  with $e_0=0.01$ fixed (left-hand side) and under the prior setting  according to the determinant\index{determinant criterion} criterion (right-hand side).}	\label{plot:trace}	\end{figure}

As reported in Section~\ref{secunkonw},  the approach by \cite{MWFSG16}
failed  when estimating the number of components for high-dimensional data with
$d=50$ as the Gibbs sampler did not converge to a smaller number of non-empty
clusters.
 When starting with an initial classification consisting of ten
(overfitting) data clusters allocated to the different components, no merging
of the components to a smaller number of  non-empty  components took place
during MCMC sampling. As can be seen  on the left-hand side of
Figure~\ref{plot:trace}, the number of observations assigned to the various
groups is constant during MCMC sampling and corresponds to the starting
classification. 
	
The main reason why Gibbs sampling  gets \lq\lq stuck\rq\rq\  lies  in the
small amount of overlapping probability between the component densities of  a
mixture model in  high-dimensional spaces.  In the starting configuration, the
data points are partitioned into many small data clusters. Due to the large
distances between component means in the high-dimensional space, the resulting
component densities are rather isolated and almost no overlapping even for
neighbouring components takes place, as can be seen in Table~\ref{tab:overlap}
where the overlapping probability for two components of the mixture model
(\ref{eq:1}) is reported.

As a consequence, in the classification step of the sampler (see Step~1 of
Algorithm~\ref{al:miy_sim}), where  an observation is assigned to the different components
according to the evaluated component densities, barely any movement can be
observed due to the missing overlap: once an observation is assigned to a
component, it is extremely unlikely that it  is allocated to another component
in the next iteration.

To overcome this curse of dimensionality,\index{curse of dimensionality} a promising  idea is to  encourage  a
priori  \lq\lq flat\rq\rq\  component densities towards achieving a stronger
overlapping of them also in higher dimensions. To this aim, we propose to
specify the prior on the component covariances $\Sigma_g \sim \Wishartinv
(c_0,C_0)$ such that a priori the ratio	of the volume  $|\Sigma_g|$ with
respect to  the total data spread $|\Cov(y)|$ is kept constant across the dimensions.
%
Using $\Ew(\Sigma_g)=C_0/(c_0-1)$, the specification of the scale matrix $C_0$
determines the prior expectation of $\Sigma_g$. In the following subsection,
guidance is given how to select $C_0$ in order to obtain a constant ratio
$|\Sigma_g|/|\Cov(y)|$.

\begin{table}[t!]
		\caption{Overlap of components two and three in mixture (\ref{eq:1}) for increasing dimension $d$}\label{tab:overlap}
                \begin{center}
		\begin{tabular}{cl}
                  \hline
			dimension $d$ & overlap \\ \hline
			$2$&  $0.034$ \\
			$4$ & $0.003$  \\
			$50$ &  $2.48\cdot 10^{-6}$\\
			$100$&  $7.34 \cdot 10^{-51}$  \\
			$200$&  $7.21\cdot 10^{-100}$ \\
                  \hline
		\end{tabular}
                \end{center}
	\end{table}

\subsection{Determinant coefficient of determination}

Consider the usual inverse Whishart prior $\Sigma_g^{-1}\sim \Wishart(c_0,C_0)$, where $c_0 =
\nu + (d - 1)/2$, with $\nu>0$ being fixed. In this subsection, we discuss the
choice of $C_0$ for high-dimensional mixtures.

We define $C_0 = \phi S_y$, with $S_y$ being the empirical covariance matrix of
the data,  as suggested in \cite{bensmail:celeux:raftery:robert:1997}, among
others,  and exploit the variance decomposition of a multivariate mixture of
normals, outlined in \citet{fruhwirth:2006}:
\begin{equation} \label{vardec}
\Cov (y)  = \sum_{g=1}^G \eta_g \Sigma_{g} +  \sum_{g=1}^G \eta_g (\mu_g-\Ew(y))(\mu_g-\Ew(y))^\top.
\end{equation}
Thus the total variance $\Cov(y)$ of a mixture distribution with $G$ components
arises from two sources of variability, namely the within-group
heterogeneity, generated by the component variances $\Sigma_{g}$, and
between-group heterogeneity, generated by the spread of the component means
$\mu_g$ around the overall mixture mean $\Ew(y)$.

To measure how much variability may be attributed to unobserved between-group heterogeneity,
\citet{fruhwirth:2006}  considers the following two coefficients of determination derived
from (\ref{vardec}), namely the trace criterion $R^2_{tr}$ and the determinant criterion $R^2_{det}$:
\begin{eqnarray}
R^2_{tr}&=& 1-\frac{\tr (\sum_{g=1}^G \eta_g\Sigma_g)}{\tr (\Cov (y) )},\label{eq:tr}\\
R^2_{det}&=&  1-\frac{|\sum_{g=1}^G \eta_g\Sigma_g|}{|\Cov (y)|}. \label{eq:det}
\end{eqnarray}
\citet[p.~170]{fruhwirth:2006}    suggests to choose $\phi$ in $C_0=\phi S_y$
such that  a certain amount of explained heterogeneity
according to the trace criterion $R^2_{tr}$ is achieved:
\begin{eqnarray}  \label{label21}
\phi_{tr}=(1-\Ew(R^2_{tr}))(c_0-(d+1)/2)\, .
\end{eqnarray}
For instance, if $c_0 = 2.5 + (d - 1)/2$, then choosing 50 percent of expected explained heterogeneity
yields $\phi_{tr} = 0.75$, while choosing a higher percentage of explained heterogeneity
such as $R^2_{tr}=2/3$ yields $\phi_{tr} = 0.5$. Obviously, the scaling factor
$\phi_{tr}$ in (\ref{label21}) is independent of the dimension $d$ of the data.
However, the  experiments  in  Section~\ref{chaptersim}  show that this
criterion yields poor clustering solutions in high dimensions such as $d=50$.

In this section, we show that choosing $\phi$ according to the determinant
criterion $R^2_{det}$ given in (\ref{eq:det})
yields a scale matrix $C_0$ in the prior of the covariance matrix $\Sigma_g$
which increases as $d$ increases. The determinant criterion $R^2_{det}$ also
measures the volume of the corresponding matrices which leads to a more
sensible choice in a model-based clustering context.

If we substitute in (\ref{eq:det}) the term $\sum\eta_g \Sigma_g$ by the prior
expected value $\Sigma_{\tilde{g}}$, with $\tilde{g}$ being an arbitrary
component, and estimate $\Cov (y)$ through $S_y$, then we obtain
\begin{eqnarray*}
R^2_{det} = 1-\frac{1}{|\Sigma^{-1}_{\tilde{g}}  S_y|} = 1 - \frac{\phi_{det}^d }{|W|},
\end{eqnarray*}
where $W \sim \Wishart  (c_0, I )$.    Taking the expectation,
\begin{eqnarray*}
\Ew(R^2_{det})=1- \phi_{det}^d \Ew(|W^{-1}|),
\end{eqnarray*}
and using
\begin{eqnarray*}
 \Ew(1/|W|^a) = \frac{\Gamma_d(c_0-a)}{\Gamma_d(c_0)},
 \end{eqnarray*}
where
\begin{eqnarray*}
\Gamma_d(c)  =  \pi^{d(d-1)/4} \prod_{j=1}^d \Gamma \left(\frac{2c+1-j}{2}\right)
\end{eqnarray*}
is the generalized Gamma function, we obtain a scaling factor $\phi_{det}$ which depends on the dimension $d$ of the data:
\begin{eqnarray}  \label{phidet}
\phi_{det}=(1- R^2_{det})^{\frac{1}{d}}\frac{\Gamma_d(c_0)}{\Gamma_d(c_0-1)}.
\end{eqnarray}
Hence, the modified prior on $\Sigma_g^{-1}$ reads:
\begin{align}\label{eq:IWnew}
\begin{aligned}
\Sigma_g^{-1}&\sim \Wishart(c_0, \phi_{det}S_y ),
\end{aligned}
\end{align}
where  $c_0=2.5+(d-1)/2$, $S_y$ is  the empirical covariance matrix of the data,   and  $\phi_{det}$ is given by  (\ref{phidet}).
Table~\ref{tab:PhiDet} reports $\phi_{det}$ for selected values of $d$ and $R^2_{det}$.
Note that for this prior $C_0=  \phi_{det}S_y$ is set to this fixed value, i.e.~no hyperprior on $C_0$ is specified.

\begin{table}[t!]
		\caption{$\phi_{det}$ for selected values of $d$ and $R^2_{det}$ with $c_0=2.5+(d-1)/2$}\label{tab:PhiDet}
	\begin{center}
          \begin{tabular}{ccccccc}
            \hline
				& $d=2$ & $d=4$ &$d=50$  &$d=100$ & $d=150$ & $d=200$ \\
			\hline
			$R^2_{det}=0.50$	&1.225  &1.831  & 11.09  & 20.55 & 29.90 & 39.22 \\
			$R^2_{det}=0.67$	&0.995  &1.651  & 11.00  & 20.46 & 29.82 & 39.14 \\
			$R^2_{det}=0.75$	&0.866  &1.540  & 10.94  & 20.41 & 29.77 & 39.08 \\
			$R^2_{det}=0.90$	&0.548  &1.225  & 10.74  & 20.22 & 29.59 & 38.90 \\
            \hline
		\end{tabular}
	\end{center}
\end{table}

%
%
%
%
%
%
%

\subsection{Simulation study using the determinant criterion}

 In order to evaluate whether the proposed determinant criterion for selecting  $C_0$ enables the Gibbs sampler to converge to the true number of components,	 a simulation study is performed. 100 data sets are sampled from the three-component mixture given in (\ref{eq:1}), with  dimensionality $d$ varying from $2$ to $200$.

	A sparse finite mixture model  is fitted to the data sets using the
Gibbs sampler and priors as described in \cite{MWFSG16}, however,  the
following prior modifications are made.  The scale parameter of the Wishart
prior on $\Sigma_k^{-1}$ is set to  $C_0=\phi_{det}  S_y$ as in
(\ref{eq:IWnew}) where $\phi_{det}$ is selected according (\ref{phidet}) with
$ R_{det}^2=0.5$, see also the first row in Table~\ref{tab:PhiDet}. This
corresponds to  a prior proportion of heterogeneity explained by the component
means of $R^2 \approx 0.67$.  The Dirichlet parameter $e_0$ is fixed to $0.01$
and we increase the maximum number $G$ of components to $G=30$.

The shrinkage factor $\lambda_l$ in the prior specification  (\ref{eq:hier}) is fixed to $1$,   since all variables $y_{il}$ are relevant for clustering in the present context. By design,  for each single variable $y_{il}$   two of the means $\mu_{l1}, \mu_{l2}, \mu_{l3}$ are different. Furthermore,  all pairs of variables
$(y_{il}, y_{i,l+2})$  exhibit three well-separated component means located at (1,1), (4,4),  (1,4) or  (4,1) in their bivariate marginal distribution $p(y_{il}, y_{i,l+2})$. Hence, none of the variables is irrelevant in the sense that  $\mu_{l1} \approx \mu_{l2} \approx \mu_{l3}$.


The estimated number of components is reported in Table~\ref{tab:results_chaper5},  based on an   Gibbs sampling with  $T=8,000$ iterations after discarding the first $800$ iterations as burn-in for each of the 100 data sets.
For $n=1,000$ and $n=10,000$,  the resulting clustering is almost perfect, even for very high dimensions such as $d=200$. The Gibbs sampler converges to the true number of components, as can be seen in the trace plot on  the right-hand side of  Figure~\ref{plot:trace}, by leaving all specified components but three empty.
	For   small  data sets with $n=100$, the criterion leads to an underfitting solution ($\hat{G}=2$) for small $d$. However, as can be seen in the left-hand scatter plot in Figure~\ref{plot:data}, a small data set of only 100 observations  may not contain enough information for estimating three groups. Also, if $d>n$ the mixture model is not well-defined, and the  Gibbs sampler gets stuck again. However, for $d=50$ and $d=100$ the approach works well also for small data sets with $n=100$.

\begin{table}[t!]
	\caption{Clustering results of the simulation study involving an  increasing number of  $d$ variables  and $n$ observations, based on 100 simulated data sets for each combination of $n$ and $d$.  The prior on $\Sigma_k^{-1}$ is chosen as in (\ref{eq:IWnew}) with $C_0=\phi_{det} S_y$ and $R^2_{det}=0.50$.
		The most frequently visited number of clusters  $\tilde{G}$   (with frequencies in parentheses), the posterior mean  of the number  of clusters $\hat{G}$ and the adjusted Rand index $\Randi $  are averaged across the 100 simulated data sets}  \label{tab:results_chaper5}
              \begin{center}
\begin{tabular}{ccccccccccc}
  \hline
		& &\multicolumn{3}{c}{$n=100$}  &\multicolumn{3}{c}{$n=1,000$}   & \multicolumn{3}{c}{$n=10,000$}  \\ 
         $d$     &            ${\phi}_{det}$                           &   $\tilde{G}$ & $\hat{G}$ &  $\Randi $   &   $\tilde{G}$ & $\hat{G}$ & $\Randi $ &   $\tilde{G}$ & $\hat{G}$ & $\Randi $     \\
		\hline
		$2$& $1.23$ & 2(93)&2.0& 0.45 & 3(100)& 3.0& 0.73 & 3(98)& 3.0& 0.75 \\
		$4$& $1.83$ & 2(70)& 2.3& 0.69 & 3(100)& 3.0& 0.93 & 3(100)& 3.0& 0.93 \\
		$50$& $11.09$ &  3(100)& 3.0& 1.00 & 3(100)& 3.0& 1.00 & 3(100)& 3.0& 1.00 \\
		$100$& $20.55$ &  3(100)& 3.0& 1.00 & 3(100)& 3.0& 1.00 & 3(100)& 3.0& 1.00 \\
		$150$& $29.90$ &  3(85)& 3.5& 0.99 & 3(100)& 3.0& 1.00 & 3(100)& 3.0& 1.00 \\
                $200$& $39.22$ &3(31)& 10.7& 0.77 & 3(97)& 3.1& 0.99 & 3(100)& 3.0& 1.00 \\
  \hline
	\end{tabular}
              \end{center}
\end{table}

\section{Concluding Remarks}	

This chapter discussed practical  Bayesian inference for finite  mixtures using
sampling based methods. While these methods work well for  moderately sized
mixtures, Gibbs sampling turned out to be the only operational method for handling
high-dimensional  mixtures. Based on a simulation study, aiming at estimating
Gaussian mixture models with fifty variables, we were unable to tune
commonly used algorithms such as the Metropolis-Hastings algorithm  or
sequential Monte Carlo,  due to the curse of dimensionality.
Gibbs sampling turned out to be the exception in this collection of algorithms.
However, we also found out that the Gibbs sampler may get stuck when
initialized with many small data clusters under previously published priors.

Hence, we consider that calibrating prior parameters in high-dimensional spaces remains
a delicate issue. For Gaussian mixtures, we examined the role of the determinant criterion
introduced by \citet{fruhwirth:2006} for incorporating the prior expected proportion of heterogeneity explained
by the component means (and thereby determining the heterogeneity covered by
the components themselves). This led  us to a new choice of the scaling matrix in
the inverse Wishart prior. The resulting prior, in combination with Gibbs sampling, worked quite well,
when estimating the number of components for Gaussian mixtures,  even for very
high-dimensional data sets with 200 variables.  Starting with a strongly overfitted
mixture, the Gibbs sampler was able to converge to the true model.
	
Many other computational issues deserve further investigation, in
particular for  mixture models  for high-dimensional data.   Additionally to
the 'big $p$' (in our notation 'big $d$') and to the  'big $n$' problem,
the 'big $G$' problem is also relevant. Are we able to estimate several tens
data clusters or more, a case that arises in high-energy physics, see \citet{fru-etal:ver} or genomics,
see \cite{RauMaugis2017}?
\index{Bayesian!inference|)}

\bibliography{CKMR}

\hyphenation{Post-Script Sprin-ger}
\begin{thebibliography}{51}
\newcommand{\enquote}[1]{``#1''}
\providecommand{\natexlab}[1]{#1}
\providecommand{\url}[1]{\texttt{#1}}
\providecommand{\urlprefix}{URL }
\expandafter\ifx\csname urlstyle\endcsname\relax
  \providecommand{\doi}[1]{doi:\discretionary{}{}{}#1}\else
  \providecommand{\doi}{doi:\discretionary{}{}{}\begingroup
  \urlstyle{rm}\Url}\fi
\providecommand{\eprint}[2][]{\url{#2}}

\bibitem[{Behboodian(1972)}]{behboodian:1972}
Behboodian J (1972).
\newblock \enquote{Information matrix for a mixture of two normal
  distributions.}
\newblock \emph{Journal of Statistical Computation and Simulation},
  \textbf{1}(4), 295--314.

\bibitem[{Bensmail \emph{et~al.}(1997)Bensmail, Celeux, Raftery, and
  Robert}]{bensmail:celeux:raftery:robert:1997}
Bensmail H, Celeux G, Raftery A, Robert C (1997).
\newblock \enquote{Inference in model-based cluster analysis.}
\newblock \emph{Statistics and Computing}, \textbf{7}(1), 1--10.

\bibitem[{Berkhof \emph{et~al.}(2003)Berkhof, {van Mechelen}, and
  Gelman}]{berkhof:mechelen:gelman:2003}
Berkhof J, {van Mechelen} I, Gelman A (2003).
\newblock \enquote{A {B}ayesian approach to the selection and testing of
  mixture models.}
\newblock \emph{Statistica Sinica}, \textbf{13}, 423--442.

\bibitem[{Bernardo and Gir{\'o}n(1988)}]{bernardo:giron:1988}
Bernardo JM, Gir{\'o}n FJ (1988).
\newblock \enquote{A {B}ayesian analysis of simple mixture problems.}
\newblock In JM~Bernardo, MH~DeGroot, DV~Lindley, AFM Smith (eds.),
  \emph{{B}ayesian Statistics 3}, pp. 67--78. Oxford University Press, Oxford.

\bibitem[{{Buchner}(2014)}]{buchner:2014}
{Buchner} J (2014).
\newblock \enquote{{A statistical test for nested sampling algorithms}.}
\newblock Preprint, arXiv:1407.5459.

\bibitem[{Capp{{\'e}} \emph{et~al.}(2004)Capp{{\'e}}, Moulines, and
  Ryd{{\'e}}n}]{cappe:moulines:ryden:2004}
Capp{{\'e}} O, Moulines E, Ryd{{\'e}}n T (2004).
\newblock \emph{Hidden {M}arkov Models}.
\newblock Springer-Verlag, New York.

\bibitem[{Celeux \emph{et~al.}(2000)Celeux, Hurn, and
  Robert}]{Celeux:Hurn:Robert:2000}
Celeux G, Hurn M, Robert C (2000).
\newblock \enquote{Computational and inferential difficulties with mixture
  posterior distributions.}
\newblock \emph{Journal of the American Statistical Association}, \textbf{95},
  957--979.

\bibitem[{Chib(1995)}]{chib:1995}
Chib S (1995).
\newblock \enquote{Marginal likelihood from the {G}ibbs output.}
\newblock \emph{Journal of the American Statistical Association}, \textbf{90},
  1313--1321.

\bibitem[{Chopin(2002)}]{chopin:2002}
Chopin N (2002).
\newblock \enquote{A sequential particle filter method for static models.}
\newblock \emph{Biometrika}, \textbf{89}, 539--552.

\bibitem[{Chopin(2004)}]{chopin:2003}
Chopin N (2004).
\newblock \enquote{Central Limit theorem for sequential {M}onte {C}arlo methods
  and its application to {B}ayesian inference.}
\newblock \emph{Annals of Statistics}, \textbf{32}(6), 2385--2411.

\bibitem[{Chopin and Robert(2010)}]{chopin:robert:2010}
Chopin N, Robert C (2010).
\newblock \enquote{Properties of nested sampling.}
\newblock \emph{Biometrika}, \textbf{97}, 741--755.

\bibitem[{Crawford \emph{et~al.}(1992)Crawford, DeGroot, Kadane, and
  Small}]{crawford:degroot:kadane:small:1992}
Crawford S, DeGroot MH, Kadane J, Small MJ (1992).
\newblock \enquote{Modelling lake-chemistry distributions: Approximate
  {B}ayesian methods for estimating a finite-mixture model.}
\newblock \emph{Technometrics}, \textbf{34}, 441--453.

\bibitem[{Del~Moral \emph{et~al.}(2006)Del~Moral, Doucet, and
  Jasra}]{delmoral:doucet:jasra:2006}
Del~Moral P, Doucet A, Jasra A (2006).
\newblock \enquote{Sequential {M}onte {C}arlo samplers.}
\newblock \emph{Journal of the Royal Statistical Society, Series B},
  \textbf{68}(3), 411--436.

\bibitem[{Dickey(1968)}]{dickey:1968}
Dickey JM (1968).
\newblock \enquote{Three multidimensional integral identities with {B}ayesian
  applications.}
\newblock \emph{Annals of Statistics}, \textbf{39}, 1615--1627.

\bibitem[{Diebolt and Robert(1990{\natexlab{a}})}]{diebolt:robert:1990b}
Diebolt J, Robert C (1990{\natexlab{a}}).
\newblock \enquote{{B}ayesian estimation of finite mixture distributions,
  {P}art {I}: {T}heoretical aspects.}
\newblock \emph{Technical Report 110}, LSTA, Universit{\'e} Paris VI, Paris.

\bibitem[{Diebolt and Robert(1990{\natexlab{b}})}]{diebolt:robert:1990a}
Diebolt J, Robert C (1990{\natexlab{b}}).
\newblock \enquote{Estimation des param{\`e}tres d'un m{\'e}lange par
  {\'e}chantillonnage bay{\'e}sien.}
\newblock \emph{Notes aux Comptes-Rendus de l'Acad{\'e}mie des Sciences I},
  \textbf{311}, 653--658.

\bibitem[{Diebolt and Robert(1994)}]{Diebolt:Robert:1994}
Diebolt J, Robert C (1994).
\newblock \enquote{Estimation of finite mixture distributions by {B}ayesian
  sampling.}
\newblock \emph{Journal of the Royal Statistical Society, Series B},
  \textbf{56}, 363--375.

\bibitem[{{Everitt} \emph{et~al.}(2016){Everitt}, {Culliford}, {Medina-Aguayo},
  and {Wilson}}]{everitt:etal:2016}
{Everitt} RG, {Culliford} R, {Medina-Aguayo} F, {Wilson} DJ (2016).
\newblock \enquote{{Sequential Bayesian inference for mixture models and the
  coalescent using sequential Monte Carlo samplers with transformations}.}
\newblock Preprint, arXiv:1612.06468.

\bibitem[{{Feroz} \emph{et~al.}(2013){Feroz}, {Hobson}, {Cameron}, and
  {Pettitt}}]{feroz:etal:2013}
{Feroz} F, {Hobson} MP, {Cameron} E, {Pettitt} AN (2013).
\newblock \enquote{{Importance nested sampling and the MultiNest algorithm}.}
\newblock Preprint, arXiv:1306.2144.

\bibitem[{Fr{\"u}hwirth \emph{et~al.}(2016)Fr{\"u}hwirth, Eckstein, and
  Fr{\"u}hwirth-Schnatter}]{fru-etal:ver}
Fr{\"u}hwirth R, Eckstein K, Fr{\"u}hwirth-Schnatter S (2016).
\newblock \enquote{Vertex finding by sparse model-based clustering.}
\newblock \emph{Journal of Physics: Conference Series}, \textbf{762}, 012055.
\newblock \urlprefix\url{http://stacks.iop.org/1742-6596/762/i=1/a=012055}.

\bibitem[{Fr\"{u}hwirth-{S}chnatter(2001)}]{fruh:2001}
Fr\"{u}hwirth-{S}chnatter S (2001).
\newblock \enquote{Markov chain {M}onte {C}arlo estimation of classical and
  dynamic switching and mixture models.}
\newblock \emph{Journal of the American Statistical Association}, \textbf{96},
  194--209.

\bibitem[{Fr{\"u}hwirth-Schnatter(2004)}]{fruhwirth:2004}
Fr{\"u}hwirth-Schnatter S (2004).
\newblock \enquote{Estimating marginal likelihoods for mixture and {M}arkov
  switching models using bridge sampling techniques.}
\newblock \emph{Econometrics Journal}, \textbf{7}(1), 143--167.

\bibitem[{Fr{\"u}hwirth-Schnatter(2006)}]{fruhwirth:2006}
Fr{\"u}hwirth-Schnatter S (2006).
\newblock \emph{Finite Mixture and Markov Switching Models}.
\newblock Springer-Verlag, New York.

\bibitem[{Fr{\"u}hwirth-Schnatter(2011)}]{fruh11}
Fr{\"u}hwirth-Schnatter S (2011).
\newblock \enquote{Dealing with label switching under model uncertainty.}
\newblock In K~Mengersen, CP~Robert, D~Titterington (eds.), \emph{Mixture
  Estimation and Applications}, chapter~10, pp. 213--239. Wiley, Chichester.

\bibitem[{Gelfand and Smith(1990)}]{gelfand:smith:1990}
Gelfand A, Smith AFM (1990).
\newblock \enquote{Sampling based approaches to calculating marginal
  densities.}
\newblock \emph{Journal of the American Statistical Association}, \textbf{85},
  398--409.

\bibitem[{Gelman and King(1990)}]{gelman:king:1990}
Gelman A, King G (1990).
\newblock \enquote{Estimating Incumbency Advantage Without Bias.}
\newblock \emph{American Journal of Political Science}, \textbf{34},
  1142{\textendash}1164.

\bibitem[{Geweke(2007)}]{geweke:2007}
Geweke J (2007).
\newblock \enquote{Interpretation and Inference in Mixture Models: Simple
  {MCMC} Works.}
\newblock \emph{Computational Statistics \& Data Analysis}, \textbf{51}(7),
  3529--3550.

\bibitem[{Green(1995)}]{green:1995}
Green P (1995).
\newblock \enquote{Reversible jump {MCMC} computation and {B}ayesian model
  determination.}
\newblock \emph{Biometrika}, \textbf{82}(4), 711--732.

\bibitem[{{Kamary} \emph{et~al.}(2018){Kamary}, {Lee}, and
  {Robert}}]{kamary:lee:robert:2015}
{Kamary} K, {Lee} JE, {Robert} CP (2018).
\newblock \enquote{{Weakly informative reparameterisations for location-scale
  mixtures}.}
\newblock \emph{Computational Statistics \& Data Analysis}.
\newblock (to appear). doi:10.1080/10618600.2018.1438900, \eprint{1601.01178}.

\bibitem[{Lee and Robert(2016)}]{lee:robert:2015}
Lee JE, Robert CP (2016).
\newblock \enquote{Importance Sampling Schemes for Evidence Approximation in
  Mixture Models.}
\newblock \emph{Bayesian Analysis}, \textbf{11}, 573--597.

\bibitem[{Lee \emph{et~al.}(2008)Lee, Marin, Mengersen, and
  Robert}]{lee:mari:meng:robe:2008}
Lee K, Marin JM, Mengersen KL, Robert C (2008).
\newblock \enquote{{B}ayesian Inference on Mixtures of Distributions.}
\newblock In NN~Sastry (ed.), \emph{Platinum Jubilee of the Indian Statistical
  Institute}. Indian Statistical Institute, Bangalore.

\bibitem[{{Malsiner-Walli} \emph{et~al.}(2016){Malsiner-Walli},
  Fr{\"u}hwirth-Schnatter, and Gr{\"u}n}]{MWFSG16}
{Malsiner-Walli} G, Fr{\"u}hwirth-Schnatter S, Gr{\"u}n B (2016).
\newblock \enquote{Model-based clustering based on sparse finite {G}aussian
  mixtures.}
\newblock \emph{Statistics and Computing}, \textbf{26}, 303--324.

\bibitem[{Marin and Robert(2010)}]{marin:robert:2010}
Marin JM, Robert CP (2010).
\newblock \enquote{Importance sampling methods for {B}ayesian discrimination
  between embedded models.}
\newblock In MH~Chen, DK~Dey, P~M{\"u}ller, D~Sun, K~Ye (eds.), \emph{Frontiers
  of Statistical Decision Making and {B}ayesian Analysis}. Springer-Verlag, New
  York.

\bibitem[{Minka(2001)}]{minka:2001}
Minka T (2001).
\newblock \enquote{Expectation Propagation for Approximate {B}ayesian
  Inference.}
\newblock In DK~Jack S~Breese (ed.), \emph{UAI '01: Proceedings of the 17th
  Conference in Uncertainty in Artificial Intelligence}, pp. 362--369.
  University of Washington, Seattle.

\bibitem[{Neal(1999)}]{neal:1999}
Neal R (1999).
\newblock \enquote{Erroneous results in ``{M}arginal likelihood from the
  {G}ibbs output''.}
\newblock \emph{Technical report}, University of Toronto.
\newblock \urlprefix\url{http://www.cs.utoronto.ca/~radford}.

\bibitem[{Rau and Maugis-Rabusseau(2018)}]{RauMaugis2017}
Rau A, Maugis-Rabusseau C (2018).
\newblock \enquote{Transformation and model choice for {RNA}-seq co-expression
  analysis.}
\newblock \emph{Briefings in Bioinformatics}, \textbf{19}, 425--436.

\bibitem[{Richardson and Green(1997)}]{richardson:green:1997}
Richardson S, Green PJ (1997).
\newblock \enquote{On {B}ayesian analysis of mixtures with an unknown number of
  components (with discussion).}
\newblock \emph{Journal of the Royal Statistical Society, Series B},
  \textbf{59}, 731--792.

\bibitem[{Robert(2007)}]{robert:2007}
Robert C (2007).
\newblock \emph{The {B}ayesian Choice}.
\newblock Springer-Verlag, New York.

\bibitem[{Robert and Casella(2004)}]{robert:casella:2004}
Robert C, Casella G (2004).
\newblock \emph{{M}onte {C}arlo Statistical Methods}.
\newblock second edition. Springer-Verlag, New York.

\bibitem[{Rousseau and Mengersen(2011)}]{rousseau:mengersen:2011}
Rousseau J, Mengersen K (2011).
\newblock \enquote{Asymptotic behaviour of the posterior distribution in
  overfitted mixture models.}
\newblock \emph{Journal of the Royal Statistical Society, Series B},
  \textbf{73}(5), 689--710.

\bibitem[{Rue \emph{et~al.}(2009)Rue, Martino, and
  Chopin}]{rue:martino:chopin:2009}
Rue H, Martino S, Chopin N (2009).
\newblock \enquote{Approximate {B}ayesian Inference for Latent {G}aussian
  Models Using Integrated Nested {L}aplace Approximations (with discussion).}
\newblock \emph{Journal of the Royal Statistical Society, Series B},
  \textbf{71}, 319--392.

\bibitem[{Skilling(2004)}]{skilling:2004}
Skilling J (2004).
\newblock \enquote{Nested Sampling.}
\newblock In R~Fisher, R~Preuss, U~{von Toussiant} (eds.), \emph{Bayesian
  Inference and Maximum Entropy Methods: 24th International Workshop}, volume
  735, pp. 395--405. AIP Conference Proceedings.

\bibitem[{Smith and Makov(1978)}]{smith:makov:1978}
Smith AFM, Makov U (1978).
\newblock \enquote{A quasi-{B}ayes sequential procedure for mixtures.}
\newblock \emph{Journal of the Royal Statistical Society, Series B},
  \textbf{40}, 106--112.

\bibitem[{Stephens(1997)}]{stephens:1997}
Stephens M (1997).
\newblock \emph{{B}ayesian Methods for Mixtures of Normal Distributions}.
\newblock Unpublished {D.Phil.} thesis, University of Oxford.

\bibitem[{Stephens(2000)}]{stephens:2000}
Stephens M (2000).
\newblock \enquote{{B}ayesian analysis of mixture models with an unknown number
  of components -- an alternative to reversible jump methods.}
\newblock \emph{Annals of Statistics}, \textbf{28}, 40--74.

\bibitem[{Tanner and Wong(1987)}]{tanner:wong:1987}
Tanner M, Wong W (1987).
\newblock \enquote{The calculation of posterior distributions by data
  augmentation.}
\newblock \emph{Journal of the American Statistical Association}, \textbf{82},
  528--550.

\bibitem[{Tierney(1994)}]{tierney:1994}
Tierney L (1994).
\newblock \enquote{{M}arkov chains for exploring posterior distributions (with
  discussion).}
\newblock \emph{Annals of Statistics}, \textbf{22}, 1701--1786.

\bibitem[{Titterington(2011)}]{titterington:2011}
Titterington D (2011).
\newblock \enquote{The {EM} Algorithm, Variational Approximations and
  Expectation Propagation for Mixtures.}
\newblock In K~Mengersen, C~Robert, D~Titterington (eds.), \emph{Mixtures:
  Estimation and Applications}, chapter~1, pp. 1--21. John Wiley, New York.

\bibitem[{Titterington \emph{et~al.}(1985)Titterington, Smith, and
  Makov}]{titterington:smith:makov:1985}
Titterington D, Smith AFM, Makov U (1985).
\newblock \emph{{S}tatistical Analysis of Finite Mixture Distributions}.
\newblock John Wiley, New York.

\bibitem[{West(1992)}]{west:1992}
West M (1992).
\newblock \enquote{Modelling with mixtures.}
\newblock In JM~Bernardo, JO~Berger, AP~Dawid, AFM Smith (eds.),
  \emph{{B}ayesian Statistics 4}, pp. 503--525. Oxford University Press,
  Oxford.

\bibitem[{Zobay(2014)}]{zobay:2014}
Zobay O (2014).
\newblock \enquote{Variational {B}ayesian inference with {G}aussian-mixture
  approximations.}
\newblock \emph{Electronic Journal of Statistics}, \textbf{8}, 355--389.
\newblock ISSN 1935-7524.
\newblock \doi{10.1214/14-EJS887}.

\end{thebibliography}
\end{document}